\documentclass[final]{statsoc}
\pdfoutput=1    

\RequirePackage[a4paper]{geometry}
\RequirePackage[OT1]{fontenc}
\RequirePackage{graphicx}
\RequirePackage{enumerate}
\RequirePackage{booktabs}
\RequirePackage{color, colortbl, transparent}
\RequirePackage{multirow}
\RequirePackage{amssymb}
\RequirePackage[cmex10]{amsmath}
\RequirePackage{placeins}
\RequirePackage{xtab}

\RequirePackage[caption=false]{subfig}
\RequirePackage{pdfpages}

\definecolor{Gray}{gray}{0.9}
\definecolor{LightGray}{gray}{0.5}

\usepackage{iproc-macros}
\newcommand{\qedhere}{}

\title[Point Process Modeling for Directed Interaction Networks]{%
    Point process modeling for directed interaction networks
}

\author[P.\ O.\ Perry]{%
    Patrick O.\ Perry
}

\address{Stern School of Business, New York University, USA}

\author[P.\ J.\ Wolfe]{%
    Patrick J.\ Wolfe
}

\address{Department of Statistical Science, University College London, UK}

\received{November 2010}
\revised{November 2012}

\runauthor{%
    P.\ O.\ Perry and P.\ J.\ Wolfe
}

\coaddress{
Patrick O.\ Perry,
Information, Operations, and Management Sciences Department,
Stern School of Business,
New York University,
44 West 4th St,
New York, NY 10012,~USA
}
\email{pperry@stern.nyu.edu}

\begin{document}


\begin{abstract}

Network data often take the form of repeated interactions between senders and
receivers tabulated over time.  A primary question to ask of such data is which traits and
behaviors are predictive of interaction.  To answer this question, a model is
introduced for treating directed interactions as a multivariate point process:
a Cox multiplicative intensity model using covariates that depend on the
history of the process.  Consistency and asymptotic normality are proved for
the resulting partial-likelihood-based estimators under suitable regularity
conditions, and an efficient fitting procedure is described.  Multicast
interactions---those involving a single sender but multiple receivers---are
treated explicitly.  The resulting inferential framework is then employed to
model message sending behavior in a corporate e-mail network.  The analysis
gives a precise quantification of which static shared traits and dynamic
network effects are predictive of message recipient selection.

\keywords{Cox proportional hazards model; Network data analysis;
Partial likelihood inference; Point processes}
\end{abstract}

\section{Introduction}
\label{S:introduction}

Much effort has been devoted to the statistical analysis of network data;
see \citet{jackson2008social}, \citet{goldenberg2009survey}, and \citet{kolaczyk2009statistical}
for recent overviews.  Often network observables comprise counts of interactions
between individuals or groups tabulated over time.  Communications networks
give rise to \emph{directed} interactions: phone calls, text messages, or
e-mails exchanged amongst a given set of individuals over a specific time
period \citep{tyler2005email,eagle2006reality}.  Specific examples of repeated
interactions from other types of networks include the following:
\citefullauthor{fowler2006connecting}'s \citeyearpar{fowler2006connecting}
study of legislators authoring and cosponsoring bills (a collaboration
network);
\citefullauthor{mckenzie2007network}'s \citeyearpar{mckenzie2007network} study
of families migrating between communities in Mexico (a migration network);
\citefullauthor{sundaresan2007network}'s \citeyearpar{sundaresan2007network}
study of zebras congregating at locations in
their habitat (an animal association network);
and \citefullauthor{papachristos2009murder}'s \citeyearpar{papachristos2009murder}
study of gangs in Chicago murdering members of rival factions (an organized
crime network).

In this article, we consider partial-likelihood-based inference for general
directed interaction data in the presence of covariates.  We first develop
asymptotic theory for the case in which interactions are strictly pairwise,
and then generalize our results to the multiple-receiver (multicast) case;
we also provide efficient algorithms for partial likelihood maximization in
these settings.  Our main assumption on the covariates is that they be
predictable, which allows them to vary with time and potentially depend on
the past.

The interaction data we consider comprise a set of triples, with triple
$(t,i,j)$ indicating that at time $t$, directed interaction
$i\rightarrow j$ took place---for instance, individual $i$ sent a message
to individual $j$.  Given such a set of triples, a primary modeling goal lies in
determining which characteristics and behaviors of the senders and
receivers are predictive of interaction.  In this vein, three important
questions stand out:

\begin{description}
    \item[Homophily] Is there evidence of homophily (an increased rate of
    interaction among similar individuals)?  To what degree is a shared
    attribute predictive of heightened interaction?

    \item[Network Effects] To what extent are past interaction behaviors predictive of future ones?
    If we observe interactions $i \to h$ and $h \to j$, are we more likely
    to see the interaction $i \to j$?

    \item[Multiplicity] How should multiple-receiver interactions of the
    type $i \rightarrow \{j_1,j_2,\ldots, j_L\}$ be modeled?  What are the
    implications of treating these as $L$ separate pairwise interactions?
\end{description}

The issues of homophily, network effects, and their interactions arise frequently in the networks literature; see, e.g., \citet{mcpherson2001birds, butts2008relational, aral2009distinguishing, snijders2010introduction}, and
references contained therein.  Multiplicity has largely been ignored in this context, however, with notable exceptions including
\citet{lunagomez2009geometric} for graphical models, and
\citet{shafiei2010mixed} for network modeling.

In the remainder of this article, we provide a modeling framework and
computationally efficient partial likelihood inference procedures to
facilitate analysis of these questions.  We employ a Cox proportional
intensity model incorporating both static and history-dependent covariates
to address the first of these two questions, and a parametric bootstrap to
address the third. Section~\ref{S:point-process-model} presents our point
process model for directed pairwise interactions, along with the resultant
inference procedures. Section~\ref{S:MPLE-consistency} establishes consistency
and asymptotic normality of the corresponding maximum partial likelihood
estimator, and Section~\ref{S:multiple-receivers} extends our framework to
the case of multiple-receiver interactions.  Section~\ref{S:enron-modeling} employs this framework to model message sending behavior in a corporate e-mail network.  Section~\ref{S:strength-of-effects} evaluates the strength of homophily and network effects in explaining these data, and Section~\ref{S:discussion} concludes the main body of the article. Appendices~\ref{S:implementation}--%
\ref{S:multiple-recipient-proofs} contain respectively implementation details and
technical results from Sections~\ref{S:MPLE-consistency}
and~\ref{S:multiple-receivers}.  The supplementary material provides comparative analyses based on related network models in the literature.

\section{A point process model and partial likelihood inference}
\label{S:point-process-model}

Every interaction process can be encoded by a multivariate counting measure.
For sender $i$, receiver $j$, and positive time $t$, define
\[
    N_t(i,j)
        =
        \#\{
            \text{directed interactions $i\rightarrow j$ in time interval
            $[0,t]$}
        \}.
\]
For technical reasons, assume that $N_0(i,j) = 0$ and that $N_t(i,j)$ is
adapted to a stochastic basis of $\sigma$-algebras
$\{ \mathcal{F}_t \}_{t \geq 0}$ satisfying the usual conditions.  Then,
$N_t(i,j)$ is a local submartingale, so by the Doob-Meyer decomposition,
there exists a predictable increasing process $\Lambda_t(i,j)$, null at
zero, such that $N_t(i,j) - \Lambda_t(i,j)$ is an $\mathcal{F}_t$-local
martingale.  Under mild conditions---the most important of which is that
no two interactions happen simultaneously---there exists a predictable
continuous process $\lambda_t(i,j)$ such that
\(
    \Lambda_t(i,j) = \int_0^t \lambda_s(i,j) \, ds.
\)
(In practical applications, simultaneous events exist and are an annoyance;
\citet{efron1977efficiency} handles simultaneity through an ad-hoc
adjustment, while \citet{brostrom2002cox} adds a discrete component
to $\Lambda$.)  The process $\lambda$ is known as the stochastic intensity
of $N$.  Heuristically,
\[
    \lambda_t(i,j) \, dt
        =
        \mathbb{P}\{
            \text{interaction $i\rightarrow j$ occurs in time interval $[t,t+dt)$}
        \}.
\]
We will model $N$ through $\lambda$ using a version of the \citet{cox1972regression}
proportional intensity model.

Let $\mathcal{I}$ be a set of senders and $\mathcal{J}$ be a (not necessarily disjoint) set of receivers.
For each sender~$i$, let $\bar \lambda_t(i)$ be a non-negative predictable
process called the baseline intensity of sender $i$; let $\mathcal{J}_t(i)$ be
a predictable finite subset of $\mathcal{J}$ called the receiver set of sender $i$.
For each sender-receiver pair $(i,j)$, let $x_t(i,j)$ be a predictable
locally bounded vector of covariates in $\reals^p$.  Let $\beta_0$
be an unknown vector of coefficients in  $\reals^p$.  For the remainder of
this section, assume that each interaction has a single receiver.

Given a multivariate counting process $N$ on
$\reals_+ \times \mathcal{I} \times \mathcal{J}$,
we model its stochastic intensity as
\begin{equation}\label{E:cox-intensity}
    \lambda_t(i,j)
        =
        \bar \lambda_t(i)
        \cdot
        \exp\{ \beta_0^\trans x_t(i, j) \}
        \cdot
        1{\{j \in \mathcal{J}_t(i)\}}.
\end{equation}
This model posits that sender $i$ in $\mathcal{I}$ interacts  with receiver $j$
in $\mathcal{J}_t(i)$ at a baseline rate $\bar \lambda_t(i)$ modulated up or
down according to the pair's covariate vector, $x_t(i,j)$.  As
\citet{efron1977efficiency} notes, the specific parametric form for the multiplier
$\exp\{ \beta_0^\trans x_t(i,j) \}$ is not central to the theoretical
analysis, but this choice is amenable to computation and gives
the parameter vector $\beta_0$ a straightforward interpretation.
\citet{butts2008relational}, \citet{vu2011continuous}, and
\citet{vu2011dynamic} used variants of this model to analyze
repeated directed actions within social settings.

The form of~\eqref{E:cox-intensity} is deceptively simple but remains
flexible enough to be useful in practice.  The model allows for
homophily and group level effects via inclusion of covariates of the form
``$1\{\text{$i$ and $j$ belong to the same group}\}$,'' where ``group'' is
some observable trait like ethnicity, gender, or age group.  Its real
strength, though, is that $x_t(i,j)$ is allowed to be \emph{any}
predictable process, in particular $x_t(i,j)$ can depend on the history
of interactions.  To model reciprocation and transitivity in the
interactions  (with $\mathcal{I} = \mathcal{J}$), for example, choose appropriate
values for $\Delta_k$ and include relevant covariates in $x_t(i,j)$:
\[
    1\{\text{interaction $j \to i$ occurred in $[t - \Delta_k,t)$}\}
\]
and
\[
    1\{\text{for some $h$, interactions $i\to h$ and $h \to j$ occurred in
             $[t - \Delta_k, t)$}\}.
\]
Any process measurable with respect to the predictable $\sigma$-algebra is
a valid covariate; this excludes only covariates depending on the future
or the immediate present.  In
Section~\ref{S:enron-covariates} we detail specific covariates suitable for
measuring homophily and network effects.

Also note that despite presuming $\mathcal{I}$ and $\mathcal{J}$ to be fixed,
our analysis
allows senders and receivers to enter and leave the study during the
observation period.  The effective number of senders at time $t$ is the
set of $i$ such that $\bar \lambda_t(i) \neq 0$, which potentially varies
with time.  Likewise, the effective number of receivers is controlled through
$\mathcal{J}_t(i)$.

Following \citet{cox1975partial}, we treat the baseline rate
$\bar \lambda_t(i)$ as a nuisance parameter and estimate the coefficient
vector $\beta_0$ using a partial likelihood.  Specifically, let
$(t_1, i_1, j_1), \ldots, (t_n, i_n, j_n)$ be the sequence of observed
interactions.  The inference procedure is motivated by decomposing
the full likelihood, $L$, as
\begin{align*}
    \begin{split}
    L(t_1&, i_1, j_1, t_2, i_2, j_2, \ldots, t_n, i_n, j_n) \\
        &=
            L(t_1, i_1)
            \, L(j_1 | t_1, i_1)
            \, L(t_2, i_2 | t_1, i_1, j_1)
            \, L(j_2 | t_2, i_2, t_1, i_1, j_1) \\
        &\quad \cdots
            L(t_n, i_n | t_{n-1}, i_{n-1}, j_{n-1}, \ldots t_1, i_1, j_1)
            \, L(j_n | t_n, i_n, t_{n-1}, i_{n-1} \ldots t_1, i_1, j_1)
    \end{split} \\
    \begin{split}
        &=
            \Big[
                L(t_1, i_1)
                \, L(t_2, i_2 | t_1, i_1, j_1)
                \cdots
                L(t_n, i_n | t_{n-1}, i_{n-1}, j_{n-1}, \ldots t_1, i_1, j_1)
            \Big] \\
        &\quad
            \cdot
            \Big[
                L(j_1 | t_1, i_1)
                \, L(j_2 | t_2, i_2, t_1, i_1, j_1)
                \cdots
                L(j_n | t_n, i_n, t_{n-1}, i_{n-1} \ldots t_1, i_1, j_1)
            \Big];
    \end{split}
\end{align*}
the term comprised of the product of conditional likelihoods of $j_1,
\dotsc, j_n$ is dubbed a partial likelihood.  In continuous time,
the log partial likelihood at time $t$, evaluated at $\beta$, is
\begin{equation}\label{E:log-pl}
    \log
    \mathit{PL}_t(\beta)
        =
        \sum_{t_m \leq t}
        \bigg\{
            \beta^\trans x_{t_m}\!(i_m, j_m)
            -
            \log\big[
                \!\!\!\!
                \sum_{j \in \mathcal{J}_{t_m}\!(i_m)}
                    \exp\{ \beta^\trans x_{t_m}\!(i_m, j)\}
            \big]
        \bigg\}.
\end{equation}
In Section~\ref{S:MPLE-consistency}, we prove under suitable regularity
conditions that the maximizer of $\log \mathit{PL}_t(\cdot)$ is a
consistent estimator of $\beta_0$ as $t$ increases.

The function $\log \mathit{PL}_t(\cdot)$ is concave, and so can be maximized
via Newton's method or a gradient-based optimization approach
\citep{nocedal2006numerical}.  These methods require one or both of the
first two derivatives of $\log\mathit{PL}_t(\cdot)$, which can be
expressed in terms of weighted means and covariances of the covariates.  The
weights are
\begin{subequations}
\begin{gather}
    w_{t}(\beta, i,j)
        =
        \exp\{ \beta^\trans x_t(i,j) \}
        \cdot
        1\{ j \in \mathcal{J}_t(i)\}, \\
    W_{t}(\beta, i)
        =
        \sum_{j \in \mathcal{J}_t(i)} w_{t}(\beta, i,j).
\end{gather}
\end{subequations}
The inner sum in $\log \mathit{PL}_t(\beta)$ is
$W_{t_m}\!(\beta, i_m)$.  The function
$\log W_{t}(\cdot, i)$ has gradient $E_{t}(\cdot, i)$ and Hessian
$V_{t}(\cdot, i)$, given by
\begin{subequations}
\begin{gather}
    E_{t}(\beta, i)
        =
        \frac{1}{W_{t}(\beta, i)}
        \sum_{j \in \mathcal{J}_t(i)}
            w_{t}(\beta, i,j) \, x_{t}(i,j), \label{E:wt-expectation}\\
    V_{t}(\beta, i)
        =
        \frac{1}{W_{t}(\beta, i)}
        \sum_{j \in \mathcal{J}_t(i)}
            w_{t}(\beta, i,j)
            \Big[ x_{t}(i,j) - E_{t}(\beta, i)\Big]^{\otimes 2},
\end{gather}
\end{subequations}
where $a^{\otimes 2} = a \otimes a = a a^\trans$.
Consequently, the gradient and negative Hessian of
$\log \mathit{PL}_t(\cdot)$ are
\begin{subequations}
\begin{gather}
    \label{E:log-pl-gradient}
    U_t(\beta)
        =
        \nabla \big[ \log \mathit{PL}_t(\beta) \big]
        =
        \sum_{t_m \leq t}
            x_{t_m}(i_m, j_m) - E_{t_m}(\beta, i_m), \\
    \label{E:log-pl-neg-hessian}
    I_t(\beta)
        =
        -\nabla^2 \big[ \log \mathit{PL}_t(\beta) \big]
        =
        \sum_{t_m \leq t}
            V_{t_m}(\beta, i_m).
\end{gather}
\end{subequations}
We call $U_t(\beta_0)$ the unnormalized score and $I_t(\beta_0)$
the observed information matrix.

Note the dependence of these terms on time-varying covariates,
which precludes using sufficient statistics and introduces
additional complexity in maximizing $\log \mathit{PL}_t(\cdot)$.
For most large interaction datasets,
existing computational routines for handling Cox models
(e.g., the function \texttt{coxph} from the \texttt{survival}
package for R \citep{therneau2009survival}) will not suffice.  In
Appendix~\ref{S:implementation}, we describe a customized method for
maximizing $\log \mathit{PL}_t(\cdot)$ that exploits sparsity in
$x_t(i,j)$.

\section{Consistency of maximum partial likelihood inference}
\label{S:MPLE-consistency}

Under the model of Section~\ref{S:point-process-model}, the maximum partial likelihood estimator (MPLE) is a natural
estimate of $\beta_0$; the inverse Hessian of $\log \mathit{PL}_t(\cdot)$
evaluated at the MPLE is a natural estimate of its covariance matrix.
We now give conditions under which these
estimators are consistent.

In the sampling regime where observation time $t$ is fixed and the number of
senders $|\mathcal{I}|$ increases, \citefullauthor{andersen1982cox}'s \citeyearpar{andersen1982cox} consistency proof for
the Cox proportional hazards model in survival analysis extends
to cover model~\eqref{E:cox-intensity}.  This setting is natural in the
context of clinical trial data, where $\mathcal{I}$ corresponds to the set of
patients under study, but does not meet the requirements typical of
interaction data.  For most interaction
data we cannot control $\mathcal{I}$ and $\mathcal{J}$, and the only way to
collect more data is to increase the observation time.  \citet{cox1972regression,cox1975partial} outlines a proof for general
MPLE consistency that applies to our sampling regime, but his argument is
heuristic; Wong's \citeyearpar{wong1986theory} treatment is more rigorous but does not cover continuous or
time-varying covariates. The general
interaction data sampling regime warrants a new consistency proof.

Our proof of consistency relies on rescaling time to make the interaction
times uniform.  To this end, define marginal processes
\(
    N_t(i) = \sum_{j \in \mathcal{J}} N_t(i,j)
\)
and
\(
    N_t = \sum_{i \in \mathcal{I}} N_t(i);
\)
also note that $t_n = \sup\{ t : N_t < n \}$ is a stopping time
and let $\mathcal{F}_{t_n}$ be the $\sigma$-algebra of events prior to
$t_n$. The main idea of the proof is to change time from the original scale
to a scale on which $t_{n} - t_{n'}$ is proportional to $n - n'$.

\subsection{Assumptions}

Let $\mathcal{B}$ be a neighborhood of $\beta_0$.  For a vector, $a$, let
$\| a \|$ denote its Euclidean norm; for a matrix, $A$, let $\| A \|$ denote
its spectral norm, equal to the largest eigenvalue of $(A^\trans A)^{1/2}$.
We require the following assumptions:
\begin{enumerate}[{A}1.]
    \item \label{A:square-int}
    \textbf{The covariates are uniformly square-integrable.}  That is,
    \[
        \E\left[
            \sup_{t, i, j} \| x_{t}(i,j) \|^2
        \right]
        \,\,\text{is bounded.}
    \]

    \item \label{A:integrated-cov-limit}
    \textbf{The integrated covariance function is well behaved.}
    When $\beta \in \mathcal{B}$ and $\alpha \in [0,1]$, as $ n \to \infty$, then with respect to the covariance function $\Sigma_\alpha(\beta)$ we have that
    \[
        \frac{1}{n}
        \sum_{i \in \mathcal{I}}
        \int_0^{t_{\lfloor \alpha n \rfloor}}
            V_s(\beta,i)
            \, W_s(\beta,i)
            \, \bar \lambda_s(i)
            \, ds
        \toP
        \Sigma_\alpha(\beta).
    \]

    \item \label{A:message-times-finite}
    \textbf{The interaction arrival times are finite.}  For each $n$,
    \[
        \mathbb{P}\{t_n < \infty\} = 1.
    \]

    \item \label{A:var-equicont}
    \textbf{The variance function is equicontinuous.}
    More precisely,
    \[
        \Big\{
            V_{t_n}(\cdot, i)
            :
            n \geq 1, i \in \mathcal{I}
        \Big\}
        \,\,\text{is an equicontinuous family of functions.}
    \]
\end{enumerate}

These technical assumptions are similar to those of
\citet{andersen1982cox}, who investigate specific settings in which their
assumptions hold.  Note that when $\| x_t(i,j) \|$ is bounded and Assumption
A\ref{A:message-times-finite} is in force, the remaining assumptions follow.

\subsection{Main results}
Assumptions A\ref{A:square-int}--A\ref{A:var-equicont} imply that the MPLE is consistent and asymptotically
Gaussian, as shown by the following two theorems.

\begin{theorem}\label{T:score-fisher}
    Let $N$ be a multivariate counting process with stochastic
    intensity as given in~\eqref{E:cox-intensity}, with true parameter
    vector $\beta_0$.  Let $t_n$ be the sequence of interaction times,
    and set $U_t(\beta)$ and $I_t(\beta)$ to
    be the gradient and negative Hessian of the log partial likelihood
    function
    as given respectively in~\eqref{E:log-pl-gradient} and~\eqref{E:log-pl-neg-hessian}.  If
    assumptions A\ref{A:square-int}--A\ref{A:integrated-cov-limit} hold, then
    as $n \to \infty$:
    \begin{enumerate}
        \item \label{I:score-part}
        $n^{-1/2} \, U_{t_{\lfloor \alpha n \rfloor}}(\beta_0)$
        converges weakly to a Gaussian process on $[0,1]$ with
        covariance function $\Sigma_\alpha(\beta_0)$;

        \item \label{I:fisher-part}
        if assumptions
        A\ref{A:message-times-finite}--A\ref{A:var-equicont} also hold, then for any consistent
        estimator $\hat \beta_n$ of $\beta_0$,
        we have that
        \[
            \sup_{\alpha \in [0,1]}
            \left\|
                \tfrac{1}{n}
                I_{t_{\lfloor \alpha n \rfloor}}(\hat \beta_{n})
                -
                \Sigma_\alpha(\beta_0)
            \right\|
            \toP
            0.
        \]
    \end{enumerate}
\end{theorem}

We don't actually require convergence of the whole sample path, but
it turns out to be just as much effort to prove as convergence of the
endpoint.  Consistency is a direct consequence of Theorem~\ref{T:score-fisher}.

\begin{theorem}\label{T:consistency}
    Let $N$ be a multivariate counting process with stochastic
    intensity as given in~\eqref{E:cox-intensity}, with true parameter
    vector $\beta_0$.  Let the log partial likelihood,
    $\log \mathit{PL}_t(\cdot)$, be as defined in~\eqref{E:log-pl}.
    Let $t_n$ be the sequence of interaction times.

    Assume that for $\beta$ in a
    neighborhood of $\beta_0$ that
    \(
        -\tfrac{1}{n} \nabla^2 [ \log \mathit{PL}_{t_n}(\beta)]
            \toP \Sigma_1(\beta),
    \)
    where $\Sigma_1(\cdot)$ is locally Lipschitz and with smallest
    eigenvalue bounded away from zero.
    If $\hat \beta_n$ maximizes $\log \mathit{PL}_{t_n}(\cdot)$ and
    conclusion~(\ref{I:score-part}) of Theorem~\ref{T:score-fisher} holds,
    then the following are true as $n\to\infty$:
    \begin{enumerate}
        \item $\hat \beta_n$ is a consistent estimator of $\beta_0$;
        \item $\sqrt{n} \, (\hat \beta_n - \beta_0)$ converges weakly
            to a mean-zero Gaussian random variable with covariance
            $[\Sigma_1(\beta_0)]^{-1}$.
    \end{enumerate}
\end{theorem}

We prove Theorems~\ref{T:score-fisher}~and~\ref{T:consistency} in Appendix~\ref{S:MPLE-consistency-proofs}.

\section{Multicast interactions}\label{S:multiple-receivers}

In Sections~\ref{S:point-process-model}~and~\ref{S:MPLE-consistency},
we have assumed that each interaction involves a single sender and a single
receiver.  The model and corresponding asymptotic theory are sufficient to
cover strictly pairwise directed interactions (e.g., phone calls), but they
do not describe interactions that can involve multiple receivers (e.g.,
e-mail messages).  We call an interaction involving a single sender and
possibly multiple receivers a multicast interaction.

In practice, multicast interactions are typically treated in an ad-hoc manner
via duplication---for example, interaction $i \to \{ j_1, j_2, j_3 \}$ gets
recorded as three separate pairwise interactions $i \to j_1$, $i \to j_2$,
and $i \to j_3$---giving rise to approximate likelihood and inference.
In this section we explore the implications of using this approximate
likelihood in the multicast setting.  In particular we show it to be closely
related to an extension of our model for directed pairwise interactions, and
that the bias introduced by such an approximation can be quantified and in
certain cases corrected.

To this end, we introduce an extension of the model to the multicast setting.
Let $\mathcal{I}$, $\mathcal{J}$, $\mathcal{J}_t(i)$, $x_t(i,j)$,
and $\beta_0$ be as in Section~\ref{S:point-process-model}.  For each sender
$i$ and positive integer $L$, let $\bar \lambda_t(i ; L)$ be a non-negative
predictable process called the baseline $L$-receiver intensity of sender $i$.
Let $(t_1, i_1, J_1), \ldots, (t_n, i_n, J_n)$ be the
sequence of observed multicast interactions, with tuple $(t, i, J)$
indicating that at  time $t$, sender $i$ interacted with receiver set $J$.
For a set $J$, let $|J|$ denote its cardinality.

Consider a model for multicast interactions where the rate of interaction
between sender $i$ and receiver set $J$ is
\begin{equation}\label{E:intensity-multiple}
    \lambda_t(i,J)
        =
        \bar \lambda_t(i ; |J|)
        \cdot
        \exp\Big\{
            \sum_{j \in J}
                \beta_0^\trans  x_t(i,j)
        \Big\}
        \cdot
        \prod_{j \in J}
        1\{ j \in \mathcal{J}_t(i) \}.
\end{equation}
The log partial likelihood at time $t$, evaluated at $\beta$, is
\begin{equation}\label{E:log-pl-multiple}
    \log
    \mathit{PL}_t(\beta)
        =
        \sum_{t_m \leq t}\!
        \bigg\{\!
            \sum_{j \in J_m}\!
                \beta^\trans x_{t_m}\!(i_m, j)
            -
            \log\big[
                \!\!\!\!
                \sum_{\substack{J \subseteq \mathcal{J}_{t_m}(i_m) \\
                               |J| = |J_m|}}\!\!\!\!\!\!\!
                    \exp\big\{
                        \sum_{j \in J}
                            \beta^\trans x_{t_m}\!(i_m, j)
                    \big\}
            \big]
        \bigg\}.
\end{equation}

Suppose instead of using the multicast model, we use duplication to get
pairwise interactions from the original multicast data.  If we use
the model of~\eqref{E:cox-intensity} for the pairwise data and ignore ties in
the interaction times, we obtain an approximate partial likelihood:
\begin{equation}\label{E:log-pl-multiple-approx}
    \log
    \widetilde{\mathit{PL}}_t(\beta)
        =\!
        \sum_{t_m \leq t}\!
        \bigg\{\!
            \sum_{j \in J_m}\!
                \beta^\trans x_{t_m}\!(i_m, j)
            -
            |J_m|
            \log\big[
                \!\!\!\!
                \sum_{j \in \mathcal{J}_{t_m}\!(i_m)}\!\!\!\!\!
                    \exp\{ \beta^\trans x_{t_m}\!(i_m, j)\}
            \big]
        \bigg\}.
\end{equation}

We claim $\log \widetilde{\mathit{PL}}_t(\beta)$ approximates
$\log \mathit{PL}_t(\beta)$.  Heuristically, replacing the
sum over all sets of size $|J_m|$ in~\eqref{E:log-pl-multiple}
with a sum over all multisets of size $|J_m|$ (i.e., allowing
duplicate elements from $\mathcal{J}_{t_m}(i_m)$), observe
\begin{align*}
    \log\big[\!\!
        \sum_{\substack{J \subseteq \mathcal{J}_{t_m}(i_m) \\
              |J| = |J_m|}}\!\!\!
            \exp\big\{
                \sum_{j \in J}
                    \beta^\trans x_{t_m}\!(i_m, j)
            \big\}
    \big]
    &\approx
        \log\big[
            \big(\!\!\!
            \sum_{j \in \mathcal{J}_{t_m}(i_m)}\!\!\!\!\!
                \exp\big\{
                    \beta^\trans x_{t_m}\!(i_m, j)
                \big\}
        \big)^{|J_m|}
        \big] \\
    &=
        |J_m|
        \log\big[\!\!\!
            \sum_{j \in \mathcal{J}_{t_m}(i_m)}\!\!\!\!\!
                \exp\big\{
                    \beta^\trans x_{t_m}\!(i_m, j)
                \big\}
        \big].
\end{align*}
In this sense,
$\log \mathit{PL}_t(\beta) \approx \log \widetilde{\mathit{PL}}_t(\beta)$.
Section~\ref{S:approximation-error} makes this statement more precise,
and Section~\ref{S:approximation-bias} analyzes the bias introduced by
maximizing $\log \widetilde{\mathit{PL}}_t(\beta)$ in lieu of
$\log \mathit{PL}_t(\beta)$.

\subsection{Approximation error}\label{S:approximation-error}
Define the receiver set growth sequence
\begin{equation}\label{E:growth-constant}
    G_n
        =
            \sum_{t_m \leq t_n}
                \frac{1\{|J_m| > 1\}}{|\mathcal{J}_{t_m}(i_m)|}.
\end{equation}
This sequence plays a critical role in bounding the error introduced by
replacing $\log \mathit{PL}$ with $\log \widetilde{\mathit{PL}}$.
Note that when $|\mathcal{J}_{t_m}(i_m)|$ is constant $G_n$ has linear
growth, but when $|\mathcal{J}_{t_m}(i_m)|$ increases, $G_n$ often has
sublinear growth.  For example, the Cauchy-Schwartz inequality gives
\[
    G_n
        \leq
            \sqrt{n}
            \cdot
            \bigg[
                \sum_{t_m \leq t_n}
                    \frac{1\{|J_m| > 1\}}{|\mathcal{J}_{t_m}(i_m)|^2}
            \bigg]^{1/2},
\]
so if $|\mathcal{J}_{t_m}(i_m)|/\sqrt{m} \to \infty$, then
$G_n = \Oh(\sqrt{n})$.  Theorem~\ref{T:log-pl-multiple-approx-error}
(proved in Appendix~\ref{S:multiple-recipient-proofs})
bounds the approximation error in terms of $G_n$.

\begin{theorem}\label{T:log-pl-multiple-approx-error}
    Let $(t_m, i_m, J_m)$ be a sequence of observations from a multivariate
    point processes with intensity as given in~\eqref{E:intensity-multiple}.
    Assume that
    \(
        \sup_t \| x_t (i,j) \|
    \)
    and
    \(
        \sup_m | J_m |
    \)
    are bounded in probability.
    If $\log \mathit{PL}$ and $\log \widetilde{\mathit{PL}}$ are as
    defined in
    \textnormal{(}\ref{E:log-pl-multiple}--\ref{E:log-pl-multiple-approx}\textnormal{)},
    and $G_n$ is as defined in~\eqref{E:growth-constant},
    then for $\beta$ in a neighborhood of $\beta_0$,
    \[
        \Big\|
        \nabla [\log \mathit{PL}_{t_n}(\beta) ]
        -
        \nabla [\log \widetilde{\mathit{PL}}_{t_n}(\beta) ]
        \Big\|
            =
            \OhP(G_n),
    \]
    and
    \[
        \Big\|
        \nabla^2 [\log \mathit{PL}_{t_n}(\beta) ]
        -
        \nabla^2 [\log \widetilde{\mathit{PL}}_{t_n}(\beta) ]
        \Big\|
            =
            \OhP(G_n).
    \]
\end{theorem}

\subsection{Bias correction from the approximate partial likelihood}
\label{S:approximation-bias}

When we use ad-hoc duplication, we are performing approximate inference
under the multicast model of~\eqref{E:intensity-multiple}.  In practice, even if we explicitly want to use the
multicast model, computing the partial likelihood of~\eqref{E:log-pl-multiple}
involves an intractable combinatorial sum, so we may resort to using the
approximation instead.  Maximizing $\log \widetilde{\mathit{PL}}_t(\cdot)$
instead of $\log \mathit{PL}_t(\cdot)$ introduces bias in the estimate of
$\beta_0$.  Theorem~\ref{T:mple-approx-error} (proved in Appendix~\ref{S:multiple-recipient-proofs}) bounds the bias.

\begin{theorem}\label{T:mple-approx-error}
    Under the setup of Theorem~\ref{T:log-pl-multiple-approx-error},
    let $\hat \beta_n$ maximize $\log \mathit{PL}_{t_n}(\cdot)$
    and let $\tilde \beta_n$ maximize
    $\log \widetilde{\mathit{PL}}_{t_n}(\cdot)$.
    Suppose for all $n$ that the Hessian
    \(
        \tfrac{1}{n}
        \nabla^2
        [ \log \mathit{\widetilde{PL}}_{t_n}(\cdot)]
    \)
    is uniformly locally Lipschitz and with smallest eigenvalue bounded away from zero
    in a neighborhood of $\hat \beta_n$.
    If $G_n/n \toP 0$, then
    \[
        \| \tilde \beta_{n} - \hat \beta_{n} \|
            =
            \OhP(G_n/n).
    \]
\end{theorem}

That $\hat \beta_n$ is a consistent estimator of
$\beta_0$ follows directly from the theory in Section~\ref{S:MPLE-consistency},
since the multicast case can be considered as a special case of
the single receiver case:  Consider the product $\mathcal{I} \times \mathbb{N}_+$
as the sender set, and the power set $\mathcal{P}(\mathcal{J})$ as the
receiver set.  For sender $(i,L)$, the process $\bar \lambda(i ; L)$ is then the
baseline send intensity, and $\{ J \subseteq \mathcal{J}_t(i) : |J| = L\}$ is the receiver set; for
sender-receiver pair $\big((i,L), J\big)$, vector $\sum_{j \in J} x_t(i,j)$ is
the covariate vector.  Consistency of the MPLE now follows from
Theorem~\ref{T:consistency}.

Suppose the true MPLE, $\hat \beta_n$, is a $\sqrt{n}$-consistent estimate of
$\beta_0$. (Theorem~\ref{T:consistency} gives sufficient conditions.)
Theorem~\ref{T:mple-approx-error} says that if
$|\mathcal{J}_{t_m}(i_m)|$ grows fast enough to make $G_n$ smaller than
$\OhP(\sqrt{n})$, then the approximate MPLE, $\tilde \beta_n$, is \emph{also}
$\sqrt{n}$-consistent.
Moreover, if $\sqrt{n}(\hat \beta_n - \beta_0)$ is asymptotically Gaussian,
then $\sqrt{n}(\tilde \beta_n - \beta_0)$ is asymptotically Gaussian with
the same covariance matrix but possibly a different mean.
Under enough regularity,
\(
    -\tfrac{1}{n} [
        \nabla^2 \log \widetilde{\mathit{PL}}_{t_n}(\tilde \beta_n)
    ]
\)
consistently estimates the limiting covariance
of $\sqrt{n}(\tilde \beta_n - \beta_0)$.  To get the mean, we use
a parametric bootstrap as follows.

Assume that the conditions of Theorem~\ref{T:mple-approx-error} hold.
The residual $\tilde \beta_n - \beta_0$ depends continuously on $\beta_0$
and the covariate process $x_t(i,j)$.  Since $\tilde \beta_n$ is a consistent
estimator of $\beta_0$, we can estimate the bias in $\tilde \beta_n$ via
a parametric bootstrap.  We generate a bootstrap replicate dataset
$\{ (t_m, i_m, J_m^{(r)}) \}$ by drawing $J_m^{(r)}$, a random subset
of $\mathcal{J}_{t_m}(i_m)$ with size $|J_m|$ whose elements are drawn
proportional to $w_{t_m}(\tilde \beta_n, i_m, \cdot)$.
We then get a bootstrap approximate MPLE, $\tilde \beta_n^{(r)}$, by
maximizing $\widetilde{\mathit{PL}}_{t_n}^{(r)}$, where
\[
    \log \widetilde{\mathit{PL}}_{t}^{(r)}\!(\beta) \\
        =
        \sum_{t_m \leq t}
        \bigg\{\!\!
            \sum_{j \in J_m^{(r)}}\!
                \beta^\trans x_{t_m}\!(i_m, j)
            -
            |J_m^{(r)}|\log\big[
                \!\!\!\!
                \sum_{j \in \mathcal{J}_{t_m}\!(i_m)}\!\!\!\!\!\!
                    \exp\{ \beta^\trans x_{t_m}\!(i_m, j)\}
            \big]
        \bigg\}.
\]
Note that $x_t(i,j)$ is determined from the original dataset, not the
bootstrap dataset.  For each bootstrap replicate, we get a residual
$\tilde \beta_n^{(r)} - \tilde \beta_n$.  With $R$ bootstrap
replicates, we estimate the bias by
\[
    \widehat{\mathrm{bias}}
        =
            \frac{1}{R} \sum_{r=1}^{R} \tilde \beta_n^{(r)} - \tilde \beta_n.
\]
We adjust for estimator bias by replacing $\tilde \beta_n$ with
$\tilde \beta_n - \widehat{\mathrm{bias}}$.

\subsection{Simulation}

We show a simulation study to empirically verify the result of
Theorem~\ref{T:mple-approx-error}.  In the study, we have one sender,
and a receiver count $|\mathcal{J}|$ ranging from $32$ to $1000$.  Each
receiver was assigned a constant covariate vector $x(j)$ whose elements were
independent Bernouli random variables with success probability $\tfrac{1}{2}$.
The components of the true coefficient vector $\beta$ were drawn independently
from the standard Normal distribution.

We chose sample sizes $n$ ranging from 32 to 100,000.  For each receiver count
$|\mathcal{J}|$, we drew $n$ multicast messages, with the receiver set $J_m$ for
message $m$ determined as follows:
we determined the size, $|J_m|$, by drawing from a geometric distribution
with success probability $p = 0.4$, so that $\mathbb{P}\{|J_m| = L\} = (1 - p)^{L -
1} \, p$ for $L \geq 1$; once $|J_m|$ was determined, we chose among
all receiver sets with cardinality $|J_m|$, with
$\mathbb{P}\{ J_m = J \} \propto \exp\{ \sum_{j \in J} \beta^\trans x(j)\}$.
Once we generated the message data, we computed $\tilde \beta$ by maximizing
the approximate log partial likelihood analogous
to~\eqref{E:log-pl-multiple-approx}.  Finally, we computed $\|\beta - \tilde
\beta\|$.

\begin{figure}[!t]
  \centering
  \makebox{\includegraphics[scale=0.55]{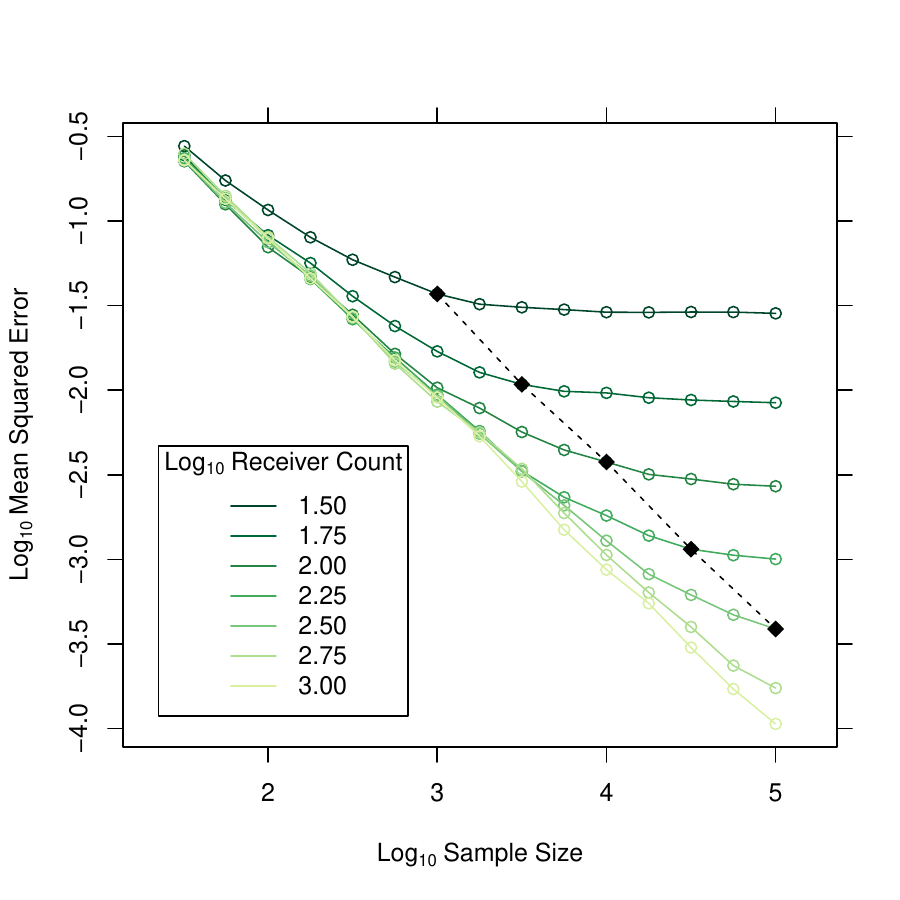}}
  \caption{
	\captiontitle{Multicast coefficient estimation error with approximate MPLE}
	Receiver count~$|\mathcal{J}|$ is equal to the square root of sample size~$n$
	along the dashed line.
  }\label{F:multicast-error}
\end{figure}

We repeated this procedure for 100 random replicates at each receiver count
and sample size, and computed the mean squared error of $\tilde \beta$ by
averaging the value of $\|\beta - \tilde \beta\|^2$ over all replicates.
Figure~\ref{F:multicast-error} displays the results.  From the spacings of the
asymptotes of the solid lines in the figure, we can see that if
$|\mathcal{J}|$ does not grow with $n$, then the error $\|\beta - \tilde
\beta\|^2$ is roughly $\Oh(|\mathcal{J}|^{-2})$ for large $n$; strictly
speaking, the assumptions of Theorem~\ref{T:mple-approx-error} do not hold in
this scenario since $G_n = \OhP(n / |\mathcal{J}|)$, but nevertheless the
theorem predicts an error rate of $\Oh(|\mathcal{J}|^{-2})$.  For the
Theorem~\ref{T:mple-approx-error} to apply, we require that $|\mathcal{J}|$
grow with $n$.  From the slope of the dashed line in
Fig.~\ref{F:multicast-error}, we can see that if $|\mathcal{J}| = \sqrt{n}$,
then $\|\beta - \tilde \beta\|^2$ is roughly $\OhP(n^{-1})$; this agrees with
the theorem, since $G_n = \sqrt{n}$ in this situation.

\section{Fitting the model to a corporate e-mail network}\label{S:enron-modeling}

Recall from Section~\ref{S:introduction} that, given a set of interaction data triples $(t,i,j)$, a primary modeling goal lies in determining which characteristics and behaviors of the senders and receivers are predictive of interaction.  The modeling and inference framework introduced above enables us to directly address these concerns, as we now demonstrate through the analysis of a corporate e-mail network consisting of a large subset of the e-mail messages sent within the Enron corporation between 1998 and 2002.  These e-mail interaction data give rise to the following questions:

\begin{description}
    \item[Homophily] To what extent are traits shared between individuals (gender, department, or seniority) predictive of interaction behaviors?

    \item[Network Effects] To what extent are dyadic or even triadic network effects, as characterized by past interaction behaviors, relevant to predicting future interaction behaviors?
\end{description}

We undertake our analysis using the multicast proportional intensity modeling framework developed in Sections~\ref{S:point-process-model} and~\ref{S:MPLE-consistency} above, employing both static covariates reflecting actor traits, as well as dynamic covariates capturing network effects.  The bootstrap technique introduced in Section~\ref{S:multiple-receivers} for multicast interactions is then used to reduce bias in the estimated effects, as well as to demonstrate that our asymptotic approximations are reasonable in this data modeling regime.  We conclude this section with a discussion of the goodness of fit of our model in this setting, before turning our attention in Section~\ref{S:strength-of-effects} to an evaluation of the strength of homophily and network effects in explaining these data.

\subsection{Data and methods}

\newcommand{\refTemployeesummary}{1} 
Our example analysis uses publicly available data from the Enron e-mail
corpus \citep{cohen2009enron}, a large subset of the e-mail messages sent
within the Enron corporation between 1998 and 2002, and made
public as the result of a subpoena by the U.S.~Federal Energy Regulatory
Commission during an investigation into fraudulent accounting practices.
We analyze the dataset compiled by \citet{zhou2007strategies}, comprising
21,635 messages sent among 156 employees between November 13, 1998 and
June 21, 2002, along with the genders, seniorities, and departments of
these employees.

Approximately 30\% of these messages have more than one recipient across
their \texttt{To}, \texttt{CC}, and \texttt{BCC} fields, with a few
messages having more than fifty recipients.
In the subsequent analysis, we exclude messages with more than 5
recipients---a subjectively-chosen cutoff that avoids e-mails sent
\emph{en masse} to large groups.

We model these data using the multicast proportional intensity
model of Section~\ref{S:multiple-receivers}, with
$\mathcal{I} = \mathcal{J} = \{ 1, 2, \ldots, 156 \}$ and
$\mathcal{J}_t(i) = \mathcal{I} \setminus \{ i \}$, and with static and
dynamic covariates described in the next section.
We fit the model by first maximizing the approximate log partial
likelihood $\log \widetilde{\mathit{PL}}_t(\beta)$ of~\eqref{E:log-pl-multiple-approx}, and then employing a
parametric bootstrap to estimate and correct the resultant bias in
parameter estimates.  We calculate standard errors using the
corresponding asymptotic theory.
In the setting of this example, the interaction count is high, so
the asymptotic framework developed in Sections~\ref{S:MPLE-consistency}
and~\ref{S:multiple-receivers} is natural.  The main violation of assumptions
A\ref{A:square-int}--A\ref{A:var-equicont} is that our covariates (described in Section~\ref{S:enron-covariates}) may
in principle be unbounded; nevertheless, bootstrap calculations
(described in Section~\ref{S:enron-bootstrap}) show that the asymptotic approximations we employ remain reasonable in this regime.

We wrote custom software in the C programming language to fit the model using
Newton's method.  Our implementation exploits structure in the covariates to
make the computational complexity of the fitting procedure roughly linear in
the number of messages and the number of actors.
Appendix~\ref{S:implementation} describes the fitting procedure in detail.  It
took approximately 20 minutes to fit the full model using a standard desktop
computer with a 2.4 GHz processor and 4GB of RAM.  Each bootstrap replicate
took approximately 10 minutes to generate and fit, using the original estimate
as a starting point for the fitting algorithm.  Most of the complexity in the
fitting procedure is due to the inclusion of triadic covariates as described
below; including only dyadic covariates reduces the fitting time to
approximately 1 minute.

\subsection{Covariates}\label{S:enron-covariates}

The goal of our investigation is to assess the predictive ability of actor traits and network effects.  To this end, we choose covariates that encode these traits and effects.  Each covariate is encoded as a component of the time-varying dyad-dependent vector $x_t(i,j)$, which is linked to the rate of interaction between sender $i$ and receiver $j$ via the multicast proportional intensity model of~\eqref{E:cox-intensity}.

\subsubsection{Static covariates to measure homophily and group-level effects}

Consider first those actor traits that do not vary with time: the actors'
genders, departments, and seniorities.  We encode the traits of actor $i$ and
their second-order interactions using 9 actor-dependent binary ($0$/$1$)
variables, as described in Fig.~\ref{F:enron-actors}.
\begin{figure}
\centering
\begin{tabular}{clc}
\toprule
Variate & Characteristic of actor $i$ & Count\\
\midrule
$L(i)$ & member of the Legal department   & 25 \\
$T(i)$ & member of the Trading department & 60 \\
$J(i)$ & seniority is Junior              & 82 \\
$F(i)$ & gender is Female                 & 43 \\
\bottomrule
\end{tabular}
\caption{
  Actor-specific traits, with counts of how many of the 156 actors share each trait
}
\label{F:enron-actors}
\end{figure}

We encode all 20 identifiable first-order interactions between the traits of
sender $i$ and receiver $j$ as components of $x_t(i,j)$.  We do this by using
variates of the form $Y(j)$ and $X(i)\cdot Y(j)$, where $X$ and $Y$ are chosen
from the list of 4 actor-dependent variates ($L$, $T$, $J$, $F$).  We also
include 4 receiver-specific covariates of the form $1 \cdot Y(j)$.
We cannot identify the coefficients for covariates of the form $X(i) \cdot 1$;
if a component of $x_t(i,j)$ is the same for all values of $j$, then the
corresponding component of $\beta$ will not be identifiable since the product
of the two can be absorbed into $\bar \lambda_t(i)$ without changing the
likelihood.

We measure homophily by way of the estimated coefficients for covariates of
the form $X(i) \cdot X(j)$.  For example, if the sum of the coefficients of
$1 \cdot J(j)$ and $J(i) \cdot J(j)$ is large and positive, this tells us that
Junior employees exhibit homophily in their choice of message recipients.

\subsubsection{Dynamic covariates to measure network effects}

Static effects are useful for determining which traits are predictive of the
relative rate of interaction between sender $i$ and receiver $j$, but they do
not shed light on network effects.  Therefore, we are also interested in the predictive
relevance of the dynamic network behaviors described in
Fig.~\ref{F:network-behaviors}.  The first two behaviors (\textbf{send} and
\textbf{receive}) are ``dyadic,'' involving exactly two actors, while the last
four (\textbf{2-send}, \textbf{2-receive}, \textbf{sibling}, and
\textbf{cosibling}) are ``triadic,'' involving exactly three actors.
\begin{figure}
\centering
\begin{tabular}{p{2cm} c p{8cm}}
  \textbf{send} &
    \vspace{1em}
    \setlength{\unitlength}{1em}
    \begin{picture}(4.75,1)
      \put(0,0){$i$}
      \put(1,0.25){\vector(1,0){2.5}}
      \put(4,0){$j$}
      \end{picture} &
    $i$ has sent $j$ a message in the past
    \\
  \textbf{receive} &
    \vspace{1em}
    \setlength{\unitlength}{1em}
    \begin{picture}(4.75,1)
      \put(0,0){$i$}
      \put(3.5,0.25){\vector(-1,0){2.5}}
      \put(4,0){$j$}
    \end{picture} &
    $i$ has received a message from $j$ in the past
    \\
  \textbf{$2$-send} &
    \vspace{1em}
    \setlength{\unitlength}{1em}
    \begin{picture}(9,1)
      \put(0,0){$i$}
      \put(1,0.25){\vector(1,0){2.5}}
      \put(4,0){$h$}
      \put(5,0.25){\vector(1,0){2.5}}
      \put(8.25,0){$j$}
    \end{picture} &
    there exists an actor $h$ such that $i$ has sent $h$ a message
    and $h$ has sent $j$ a message in the past
    \\
  \textbf{$2$-receive} &
    \vspace{-3em}
    \setlength{\unitlength}{1em}
    \begin{picture}(9,1)
      \put(0,0){$i$}
      \put(3.5,0.25){\vector(-1,0){2.5}}
      \put(4,0){$h$}
      \put(7.75,0.25){\vector(-1,0){2.5}}
      \put(8.25,0){$j$}
    \end{picture} &
    there exists an actor $h$ such that $i$ has received a message
    from $h$, and $h$ has received a message from~$j$
    \\
  \textbf{sibling} &
    \setlength{\unitlength}{1em}
    \begin{picture}(5,5)(0,4)
      \put(2.5,4){$h$}
      \put(2.5,3.5){\vector(-1,-2){1.25}}
      \put(2.75,3.5){\vector(1,-2){1.25}}
      \put(1,0){$i$}
      \put(3.75,0){$j$}
    \end{picture} &
    there exists an actor $h$ such that $h$ has sent $i$ and $j$
    messages in the past
    \\
  \textbf{cosibling} &
    \vspace{4em}
    \setlength{\unitlength}{1em}
    \begin{picture}(5,5)(0,4)
      \put(2.5,4){$h$}
      \put(1.25,1){\vector(1,2){1.25}}
      \put(4,1){\vector(-1,2){1.25}}
      \put(1,0){$i$}
      \put(3.75,0){$j$}
    \end{picture} &
    there exists an actor $h$ such that $h$ has received messages
    from $i$ and~$j$
\end{tabular}
\caption{
  Dynamic covariates to measure network effects
}
\label{F:network-behaviors}
\end{figure}

To measure first-order dependence of message exchange behavior on these network effects, we introduce binary
indicators for all $6$ effects as components of $x_t(i,j)$.  These indicators
depend on the sender $i$, the receiver, $j$, and the history of the process at
the current time $t$.  By the shorthand notation $1\{\textbf{send}\}$, we denote
the indicator variable depending on sender $i$, receiver $j$, and the current
time, $t$,  which indicates if $i$ has sent $j$ a message before time $t$, with the remaining notations ($1\{\textbf{receive}\}$, $1\{\textbf{2-receive}\}$, etc.)
defined similarly.

To measure higher-order time dependence, we introduce additional covariates of the following form.  We partition the interval $[-\infty, t)$ into $K = 7$ sub-intervals:
\[
  [-\infty, t) =
  [t - \Delta_K, t - \Delta_{K-1}) \cup [t - \Delta_{K-1}, t - \Delta_{K-2}) \cup \dotsb \cup [t - \Delta_1, t - \Delta_0)
\]
where $\infty = \Delta_K > \Delta_{K-1} > \dotsb > \Delta_1 > \Delta_0 = 0$
and ``$t - \infty$'' is defined to be $-\infty$.  Specifically, we set
$\Delta_k = (7.5\text{ minutes}) \times 4^k$ for $k = 1, \dotsc, K-1$ so that
for $k$ in this range $\Delta_k$ takes the values $30\text{ minutes}$,
$2\text{ hours}$, $8\text{ hours}$, $32\text{ hours}$, $5.33\text{ days}$, and
$21.33\text{ days}$.

Define the half-open interval $I_{t}^{(k)} = [t - \Delta_k, t -
\Delta_{k-1})$.  For $k = 1, \dotsc, K$ we define the dyadic~effects
\begin{align*}
  \text{\textbf{send}}^{(k)}_t(i,j)
    &= \#\{ i \to j \text{ in } I_t^{(k)} \}, \\
  \text{\textbf{receive}}^{(k)}_t(i,j)
    &= \#\{ j \to i \text{ in } I_t^{(k)} \};
\end{align*}
for sender $i$, such that these covariates measure the number of messages sent to, and respectively received by, receiver $j$ in time interval $I_t^{(k)}$.

The dyadic effects have been defined in the manner above to enable easy interpretation of
the corresponding coefficients.  To illustrate this, for $k = 1, \dotsc, K$,
suppose that $\beta_{k}$ is the coefficient corresponding to
$\text{\textbf{send}}^{(k)}_t(i,j)$.  If we observe the message $i \to j$ at
time $t$, then for future time $t'$ in the interval $(t, t + \Delta_1]$, the
rate $\lambda_{t'}(i,j)$ will be multiplied be the factor $e^{\beta_{1}}$; for
$t'$ in the interval $(t + \Delta_1, t + \Delta_2]$, the rate will be
multiplied by $e^{\beta_{2}}$; this continues similarly, with the rate being
multiplied by $e^{\beta_{k}}$ whenever $t' \in (t + \Delta_{k-1}, t +
\Delta_{k}]$; equivalently, when $\Delta_{k-1} < t' - t \leq \Delta_k$.  Thus,
the coefficients $\beta_{1}, \dotsc, \beta_{K}$ measure the effect of a ``send
event'' and how this effect decays over time.  We expect that $\beta_{k}$ will
decrease as $k$ increases, but we do not enforce this constraint on the
estimation procedure.

The triadic effects involve pairs of messages.  For $k = 1, \dotsc, K$ and $l = 1, \dotsc, K$ we define the triadic effects
\begin{align*}
  \text{\textbf{2-send}}^{(k,l)}_t(i,j)
    &= \sum_{h \neq i,j}
      \#\{ i \to h \text{ in } I_t^{(k)} \}
      \cdot \#\{ h \to j \text{ in } I_t^{(l)} \}, \\
  \text{\textbf{2-receive}}^{(k,l)}_t(i,j)
    &= \sum_{h \neq i,j}
      \#\{ h \to i \text{ in } I_t^{(k)} \}
      \cdot \#\{ j \to h \text{ in } I_t^{(l)} \}, \\
  \text{\textbf{sibling}}^{(k,l)}_t(i,j)
    &= \sum_{h \neq i,j}
      \#\{ h \to i \text{ in } I_t^{(k)} \}
      \cdot \#\{ h \to j \text{ in } I_t^{(l)} \}, \\
  \text{\textbf{cosibling}}^{(k,l)}_t(i,j)
    &= \sum_{h \neq i,j}
      \#\{ i \to h \text{ in } I_t^{(k)} \}
      \cdot \#\{ j \to h \text{ in } I_t^{(l)} \}.
\end{align*}
For sender $i$ and receiver $j$, the covariate
$\text{\textbf{2-send}}^{(k,l)}_t(i,j)$ counts the pairs of messages such that for some $h$ distinct from $i$ and $j$, message $i \to h$ occurred in interval $I_t^{(k)}$ and message $h \to j$ occurred in interval $I_t^{(l)}$; the other covariates behave similarly.

As with the dyadic effects, the triadic effects are designed so that their
coefficients have a straightforward interpretation.  However, since triadic
effects involve pairs of messages, the interpretation is a bit more involved.
We illustrate with the $\text{\textbf{2-send}}^{(k,l)}_t(i,j)$ covariate
having coefficient $\beta_{k,l}$ for $k = 1, \dotsc, K$ and $l = 1, \dotsc,
K$.  Take $i$ and $j$ to be two actors.  Suppose at time $t$ we observe the
message $h \to j$.  At this point, we look through the history of the process
for all messages of the form $i \to h$; when paired with the original $h \to
j$ message, each of these defines a ``2-send event.''  The other 2-send events
are defined as follows: if at time $s$ we observe the message $i \to h$, then
we enumerate all observed messages $h \to j$ in the history of the process;
when each of these is paired with the original $i \to h$ event it constitutes
a 2-send event.  A pair $(s,t)$ can be associated with each 2-send event,
where $s$ is the time of the $i \to h$ message and $t$ is the time of the $h
\to j$ message.  At time $t'$ after $s$ and $t$, the existence of the 2-send
event causes the sending rate $\lambda_{t'}(i,j)$ to be multiplied by the
factor $e^{\beta_{k,l}}$, where $t' \in (s + \Delta_{k-1}, s + \Delta_{k}]$
and $t' \in (t + \Delta_{l-1}, t + \Delta_l]$.  We expect $\beta_{k,l}$ to
decrease as $k$ and $l$ increase, though again we do not enforce this
constraint in the fitting procedure.

As previously noted, \cite{butts2008relational} used a variant of the proportional
intensity model to capture interaction behavior in social settings.  As such, a correspondence can be drawn between certain of the covariates in \cite{butts2008relational} and those outlined above.  If we set
$K = 1$, then the $\textbf{send}_t$ covariate is equivalent to an unnormalized version
of Butts' persistence covariate, and the sum $(\textbf{send}_t +
\textbf{receive}_t)$ becomes an unnormalized version of Butts' preferential
attachment covariate.  For the triadic effects, Butts' OTP, ITP, ISP, and OSP
covariates are analogous to the $\textbf{2-send}$, $\textbf{2-receive}$,
$\textbf{sibling}$, and $\textbf{cosibling}$ covariates, although the exact definitions differ slightly.
(For example, $\textrm{OTP}_t(i,j)$ is
defined as $\sum_{h} \min[ \#\{ i \to h \text{ in } (-\infty, t) \}, \,
\#\{ h \to j \text{ in } (-\infty, t) \} ]$.)  The versions of these covariates that we have introduced above, however, are designed to enable a more precise quantification of the time-dependence of network effects, as well as a more straightforward interpretation of the corresponding coefficients.  In related models, \cite{vu2011continuous, vu2011dynamic} use similar
covariates, except that they do not partition $[-\infty, t)$
into sub-intervals.

\subsection{Bootstrap bias correction}\label{S:enron-bootstrap}

Given the model specification, data, and covariates outlined above, we can
estimate the parameter vector $\beta_0$ under the approximate
log partial likelihood of~\eqref{E:log-pl-multiple-approx}.  Recall
that the results of Section~\ref{S:multiple-receivers} bound the bias
resulting from this approximate MPLE procedure as a function of the growth
rate of the recipient set $\mathcal{J}$ over time.  Here, treating the
set $\mathcal{J}$ of 156 Enron employees as constant, the resultant bias
is of order $1/|\mathcal{J}|$---and, since $|\mathcal{J}| = 156$ is on the
order of the square root of the number 21,365 of messages in the dataset,
we can correct this bias using the parametric bootstrap outlined at the
end of Section~\ref{S:multiple-receivers}.

\begin{figure}
    \centering
    \makebox{\includegraphics[scale=0.67]{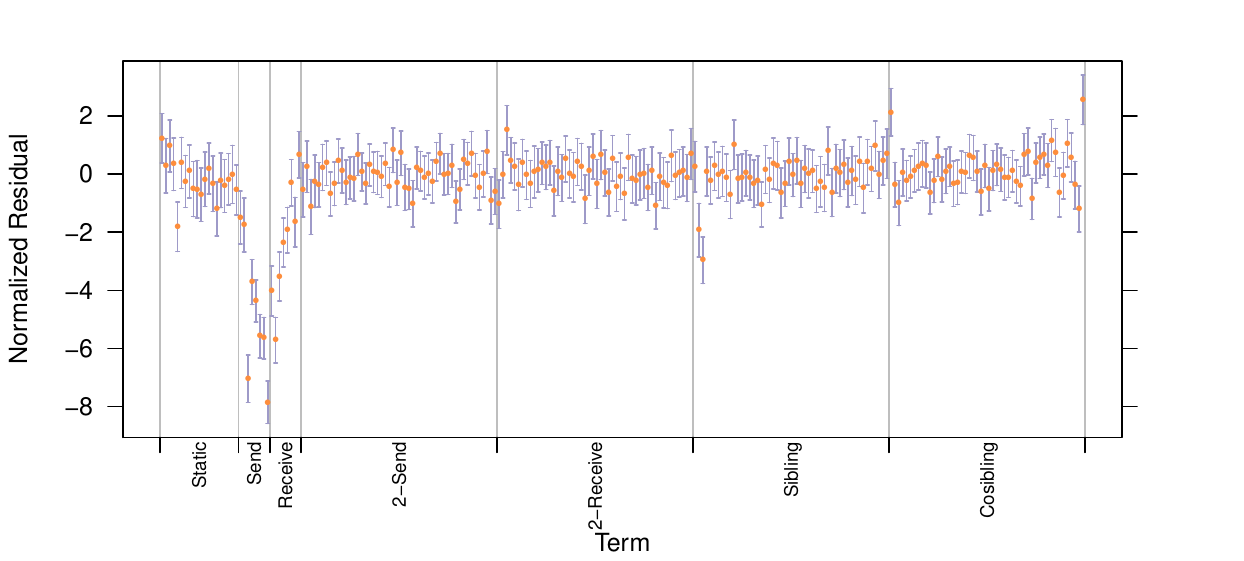}}
    \caption{
        \captiontitle{Enron bootstrap residuals}
        Summary of bootstrap residuals
        for estimated coefficients using the Enron dataset, normalized by
        estimated standard errors.  The points (orange) show the means, and
        the error bars (purple) show one standard deviation.  Coefficients
        are grouped by model term.
    }
    \label{F:boot-resid}
\end{figure}

Fig.~\ref{F:boot-resid} summarizes the corresponding bootstrap residuals (from
$500$ replicates) for each component of the estimated parameter vector
$\beta_0$; we can see from this figure that treating messages with multiple
recipients as multiple single-recipient messages introduces bias on the order of the
standard error for most of the coefficients.  There is a pronounced negative
bias in coefficient estimates for the dyadic effects, which is representative
of a more general phenomenon.  Sparsity in the components of $x_{t}(i,j)$
(when considered as a function of $j$), when combined with high values of the
corresponding entries $\beta$, leads to negative bias in the coefficient
estimates when there are messages with multiple recipients.  The approximation
in~\eqref{E:log-pl-multiple} is worst when for some $j^\ast$, weight
$w_{t_m}(i_m,j^\ast)$ far exceeds all other values of $w_{t_m}(i_m,j)$, so
that $w_{t_m}(i_m, j^\ast) \approx W_{t_m}(i_m)$; when $|J_m|$ is large, the
maximum of $\widetilde{\mathrm{PL}}$ will avoid this situation by shrinking
$\beta$ where $x_{t_m}(i_m, j)$ is sparse.  The dyadic covariates are
particularly sparse, so the estimates for their coefficients are particularly
vulnerable to this bias.

Besides correcting for bias, the bootstrap simulations give us confidence that
the asymptotic approximations are reasonable.  The simulated standard errors
are very close to those predicted by the theory, despite the norm $\| x_t(i,j)
\|_2$ being potentially unbounded, contrary to the assumptions of
Theorem~\ref{T:score-fisher}.

\subsection{Goodness of fit}\label{S:enron-goodness}

Figure~\ref{F:deviance} details an ad-hoc analysis of deviance for the fitted
model, showing how the approximate deviance (twice the approximate log partial
likelihood) behaves as we add consecutive terms to the model.  Group-level
(static) effects account for 15\% of the null deviance and network effects
account for 37\%.  The most dramatic decrease in the residual deviance comes
from introducing the ``Send'' terms into the model; with only 8 degrees of
freedom, they are able to account for 33\% of the null deviance.  The full
model accounts for $52\%$ of the null deviance.

\begin{figure}
\centering
\begin{tabular}{lrrrr}
    \toprule
    \textbf{Term}
        & \textbf{Df}
        & \textbf{Deviance}
        & \textbf{Resid. Df}
        & \textbf{Resid. Dev} \\
    \midrule
    Null      &    &        & 32261 & 325412 \\
    Static    & 20 &  50365 & 32241 & 275047 \\
    Send      &  8 & 107942 & 32233 & 167105 \\
    Receive   &  8 &   5919 & 32225 & 161186 \\
    Sibling   & 50 &   3601 & 32175 & 157585 \\
    2-Send    & 50 &    516 & 32125 & 157069 \\
    Cosibling & 50 &   1641 & 32075 & 155428 \\
    2-Receive & 50 &    158 & 32025 & 155270 \\
    \bottomrule
\end{tabular}
    \caption{
        Ad-hoc analysis of deviance for the Enron model.  Residual deviance
        is defined as twice the approximate negative log partial likelihood
        from~\eqref{E:log-pl-multiple-approx}.
        The ``Static'' term contains the group level effects, and the
        other terms contain the network effects.
    }
    \label{F:deviance}
\end{figure}

The residual deviance for the full model is approximately $4.8$ times the
residual degrees of freedom, and so an ad-hoc adjustment for this over-dispersion
is to multiply the calculated standard errors by $\sqrt{4.8} \approx 2.2$.

Note, however, that the residual deviance by itself is not adequate as a
goodness-of-fit measure, as it depends only on the estimated coefficients (see
Section~4.4.5 of \cite{mccullagh1989generalized} for discussion of a related
problem for logistic regression with sparse data).  To shed more light on how
well the model fits these data, we use a normalized version of the martingale
residual from \cite{therneau1990martingale}, which we call a Pearson residual.
Specifically, given $\hat \beta$, we define
\[
  \hat N_{t}(i,j)
    = \sum_{t_m \leq t}
        \frac{w_{t_m}(\hat \beta, i, j)}{W_{t_m}(\hat \beta, i)}
        1\{i_m = i\},
\]
which is the expected number of $i \to j$ events given the estimated model,
with $\int \bar \lambda_t(i) \, dt$ estimated by the
\cite{breslow1974covariance} estimate $\int W_t(\hat \beta, i)^{-1} \sum_j dN_{i,j}(t)$.  The martingale residual analogous to that of \cite{therneau1990martingale} is then
defined as $N_t(i,j) - \hat N_t(i,j)$; we normalize this quantity by an
estimate of its standard deviation to get a ``Pearson'' residual:
\(
    {(N_t(i,j) - \hat N_t(i,j))}/\{\hat N_t(i,j)\}^{1/2}.
\)

\begin{figure}[!t]
  \centering
  \subfloat[
    Observed count $N_{\infty}(i,j)$ plotted against
    expected count $\hat N_{\infty}(i,j)$
  ]{\label{F:nobs-by-nexp}
    \makebox{\includegraphics[scale=0.59]{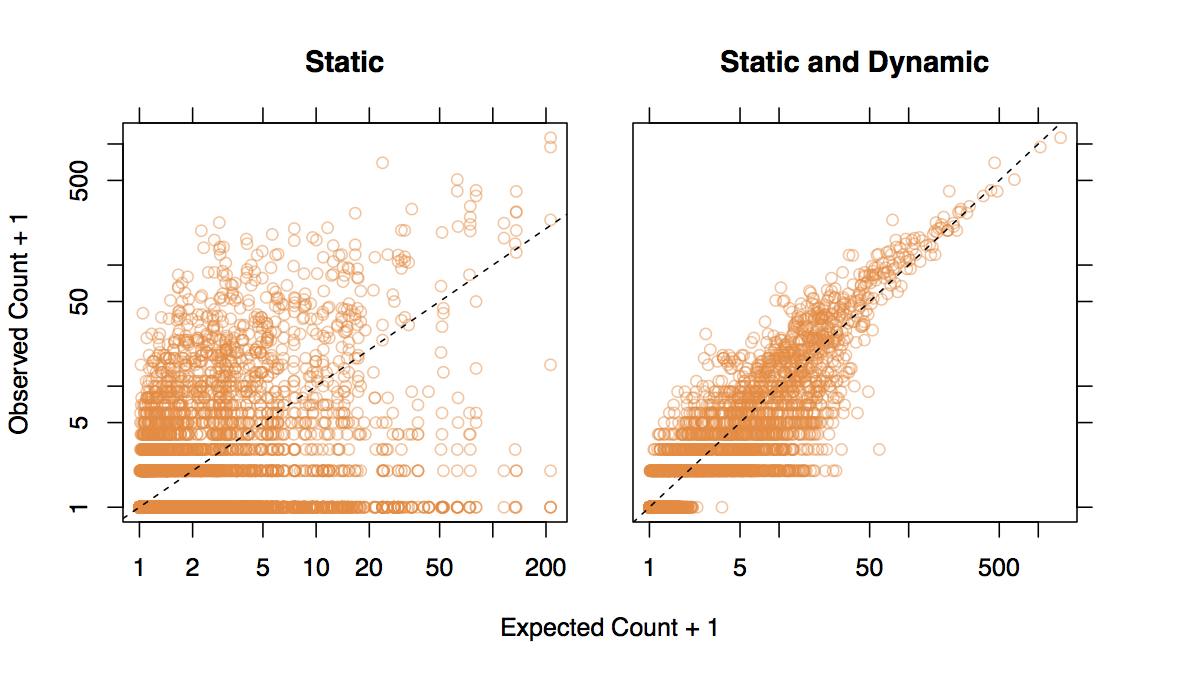}}}
  \\
  \subfloat[
    Pearson residual
    $(N_\infty(i,j) - \hat N_\infty(i,j)) / \{\hat N_\infty(i,j)\}^{1/2}$
    vs.~expected count
  ]{\label{F:resid-by-nexp}
    \makebox{\includegraphics[scale=0.59]{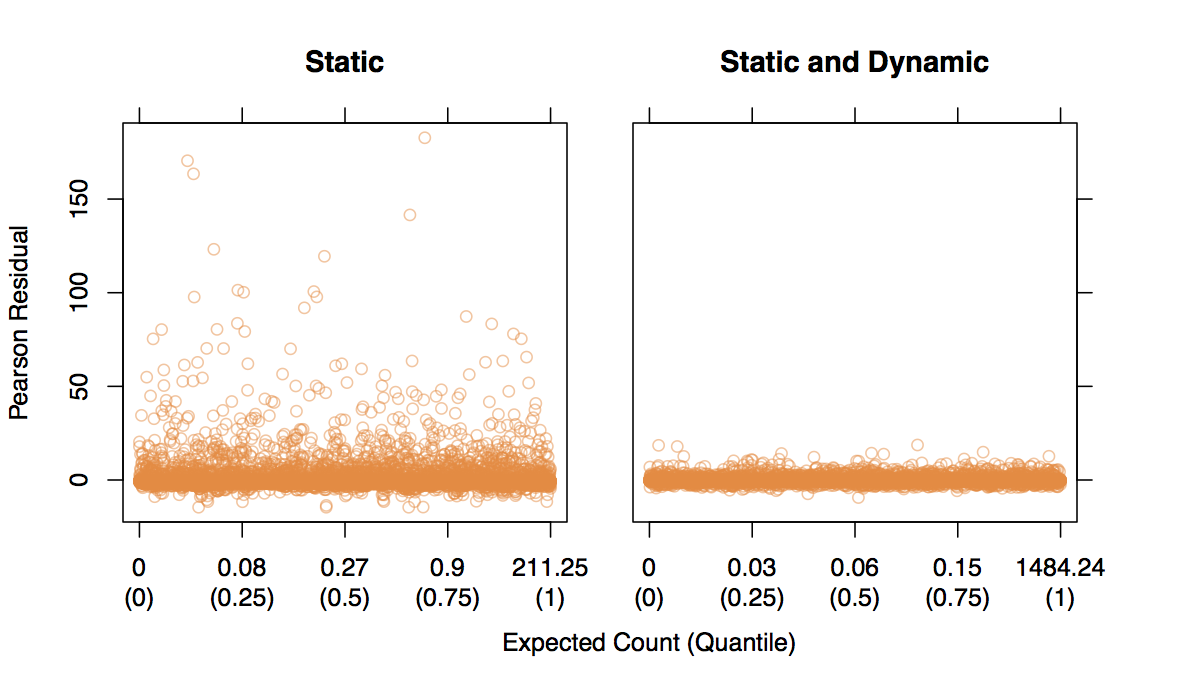}}}
  \caption{\label{F:goodness}
    Goodness of fit plots for two models
  }
\end{figure}

Fig.~\ref{F:nobs-by-nexp} shows a plot of $N_\infty(i,j)$ versus $\hat
N_\infty(i,j)$ for two different models.  In the ``static'' model, we only
include the static covariates, while in the full (``static and dynamic'') model,
we also include all six types of network covariates.  The fit for the static model
is poor.  For instance, it repeatedly predicts up to 200 $i \to
j$ events where we only observed 1 or 2; likewise, the model predicts $1$ or
fewer events where we observed up to $20$.  For the full model, which includes
the dynamic covariates to account for network effects, the fit is much better, with the relationship between
observed and expected interaction counts being roughly linear.

Fig.~\ref{F:resid-by-nexp} shows the Pearson residuals.  For the full model,
more than 95\% are less than $1.21$ in absolute value, and the maximum
absolute residual is $18.7$.  In contrast, the 95\% quantile for the absolute
residuals in the static model is at $3.5$, and the maximum absolute residual
is $182.7$.  The sum of squares if the residuals ($X^2$) is $17281$ in the
full model, over $34$ times lower than that for the static model ($596253$).
We don't know what a ``reasonable'' value for $X^2$ is; an ad-hoc degrees of
freedom calculation suggests that for the full number this should be roughly
equal to $23944 = 156 \cdot 155 - (20 + 2 \cdot 8 + 4 \cdot 50)$, which
suggests that the full model is too aggressive.  The bootstrap simulations
confirm this, with $17055$ being $5.6$ standard deviances below the mean
value $X^2$ for the bootstrap replicates.

For a more parsimonious model, we might drop most of the triadic effects.
Indeed, the model which only uses dyadic effects has a $X^2$ value of $21094$.
However, at this stage we desire a model with the lowest possible bias, and
also wish to acquire estimates for all of the network effects.

\section{Evaluating the strength of homophily and network effects}\label{S:strength-of-effects}

Given the model fitting procedure and results described above, we may now
evaluate the strength of homophily and network effects in predicting the
interaction behavior observed in our data.

\subsection{Assessing evidence for homophily in the Enron data}\label{S:enron-homophily}

The analyses of Section~\ref{S:enron-modeling} above have established that our
multicast proportional intensity model with chosen covariates is reasonably
accurate in describing message recipient selection, conditional on the sender
and the history of the process.  Thus, we are justified in using the estimated
coefficients from the model to assess the predictive ability of the
corresponding covariates.

Our first task is to gauge the predictive strength of homophily.  To this end,
Fig.~\ref{T:group-dynamic} shows the estimated group-level coefficients for
our model.  Notably, homophily is evident for all almost all main effects
(Department, Seniority, and Gender): the estimated coefficients of $L(j)$,
$T(j)$, and $J(j)$ are all negative, while the sum of the estimated
coefficients of $F(j)$ and $F(i) \cdot F(j)$ is positive.
Negative homophily is evidenced in that the sum of the coefficients for
$L(j)$ and $L(i) \cdot L(j)$ is negative.
The coefficient of $F(j)$ and the sum of the coefficients for
$T(j)$ and $T(i) \cdot T(j)$; and
$J(j)$ and $J(i)\cdot J(j)$ are not significant.

Taking Gender as an example, the way the homophily effect manifests is as
follows: if $i$ is a Female sending a message at time $t$, and person $j$ is
identical to person $j'$ except for Gender, then $i$ is more likely to send to
the similarly-gendered individual.  The relative rate is $\exp(0.04 + 0.15)
\approx 1.2$.  The characterization for other types of homophily is similar.

Conspicuously, the only example of negative homophily is when the sender $i$ is in
the Legal department.  In this case, if person $j$ is identical to person $j'$
except for Department, then $i$ is more likely to send to an individual in a
different department.  The relative rates for the three departments are
$\exp(0.63 - 0.91) \approx 0.76$ for the Legal department, $\exp(0.28-0.36) \approx
0.92$ for the Trading department, and $\exp(0) = 1$ for any Other department.

\begin{figure}
  \centering
  \makebox[\textwidth]{
    \begin{tabular}{lrrrr}
\toprule
& \multicolumn{4}{c}{Receiver} \\
\cmidrule(l){2-5}
Sender & \multicolumn{1}{c}{L} & \multicolumn{1}{c}{T} & \multicolumn{1}{c}{J} & \multicolumn{1}{c}{F} \\
\midrule
\multirow{2}{*}{1} &-0.91 &-0.36 &-0.34 &\textcolor{LightGray}{0.04}\\
 &\scriptsize{(0.04)} &\scriptsize{(0.04)} &\scriptsize{(0.04)} &\textcolor{LightGray}{\scriptsize{(0.03)}}\\[1ex]
\multirow{2}{*}{L} &\cellcolor{Gray}0.63 &0.28 &0.22 &0.15\\
 &\cellcolor{Gray}\scriptsize{(0.05)} &\scriptsize{(0.05)} &\scriptsize{(0.04)} &\scriptsize{(0.04)}\\[1ex]
\multirow{2}{*}{T} &0.32 &\cellcolor{Gray}0.43 &0.27 &\textcolor{LightGray}{-0.07}\\
 &\scriptsize{(0.07)} &\cellcolor{Gray}\scriptsize{(0.05)} &\scriptsize{(0.05)} &\textcolor{LightGray}{\scriptsize{(0.05)}}\\[1ex]
\multirow{2}{*}{J} &\textcolor{LightGray}{0.06} &0.28 &\cellcolor{Gray}0.37 &-0.13\\
 &\textcolor{LightGray}{\scriptsize{(0.05)}} &\scriptsize{(0.04)} &\cellcolor{Gray}\scriptsize{(0.03)} &\scriptsize{(0.03)}\\[1ex]
\multirow{2}{*}{F} &0.59 &-0.21 &\textcolor{LightGray}{-0.09} &\cellcolor{Gray}0.15\\
 &\scriptsize{(0.05)} &\scriptsize{(0.05)} &\textcolor{LightGray}{\scriptsize{(0.04)}} &\cellcolor{Gray}\scriptsize{(0.03)}\\[1ex]
\bottomrule
\end{tabular}

  }
  \caption{
    Estimated coefficients and standard errors for group-level covariates
    of the form $X(i) \cdot Y(j)$, where $i$ is the sender, $j$ is the
    receiver, and $X(i)$ and $Y(j)$ are given in the row and column
    headings; dark coefficients are significant (via Wald test) at
    level $10^{-3}$.
  }
  \label{T:group-dynamic}
\end{figure}

Were we interested only in homophily, we might be tempted to forgo
the proportional intensity model of~\eqref{E:cox-intensity}, and instead
perform a contingency table analysis.  
The supplementary material
explores this approach in detail.  The major shortcoming of the contingency
table approach is that it assumes that the messages are independent, which
leads to bias in the parameter estimates.

\subsection{Evaluating the importance of network effects}
\label{S:enron-network}

In Section~\ref{S:enron-homophily} we established that homophily was
predictive of sending behavior, even after accounting for network effects.  We
now investigate the characteristics of these network effects and
establish which of these effects are of greatest importance.

To begin our analysis, Fig.~\ref{F:enron-net-indicator} shows the estimated coefficients for the
network indicator effects, giving a crude picture of the predictive importance
of each network effect.  The estimated coefficients are all positive,
indicating that network effects strengthen the ties between individuals.  The
estimated coefficient for $1\{\textbf{send}\}$ is over three times larger
than the other coefficients, agreeing with the general notion that one is most likely
to do today the things one did yesterday.  The next tier of indicator effects
comprises $1\{\textbf{receive}\}$, $1\{\textbf{sibling}\}$, and
$1\{\textbf{2-send}\}$, whose estimated coefficients range from $0.67$ to $1.06$.
Two triadic effects, $1\{\textbf{2-receive}\}$ and
$1\{\textbf{cosibling}\}$, are not significantly predictive of sending behavior.

\begin{figure}
\centering
\footnotesize
\begin{tabular}{lrrrrrr}
\toprule
Variate
& $1\{\textbf{send}\}$
& $1\{\textbf{receive}\}$
& $1\{\textbf{2-send}\}$
& $1\{\textbf{2-receive}\}$
& $1\{\textbf{sibling}\}$
& $1\{\textbf{cosibling}\}$
\\
\midrule
Coefficient
& 3.26
& 0.97
& 0.67
& \textcolor{LightGray}{0.01}
& 1.06
& \textcolor{LightGray}{0.09}
\\
(SE)
& \scriptsize{(0.03)}
& \scriptsize{(0.02)}
& \scriptsize{(0.05)}
& \textcolor{LightGray}{\scriptsize{(0.04)}}
& \scriptsize{(0.05)}
& \textcolor{LightGray}{\scriptsize{(0.04)}}
\\
\bottomrule
\end{tabular}
\normalsize
\caption{Estimated coefficients for network indicator effects}
\label{F:enron-net-indicator}
\end{figure}

\begin{figure}
  \centering
  \makebox{\includegraphics[scale=0.62]{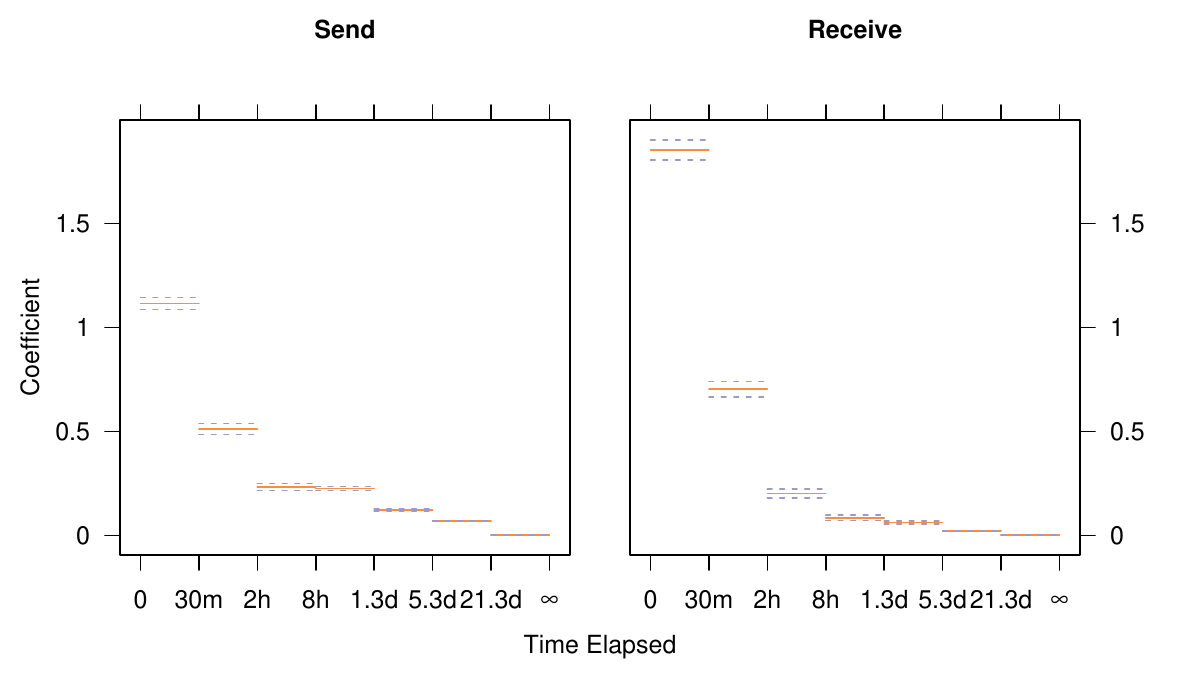}}
  \caption{Estimated coefficients for dyadic effects, with standard errors}
  \label{F:enron-dyad}
\end{figure}

\begin{figure}
  \centering
  \makebox{\includegraphics[scale=0.62]{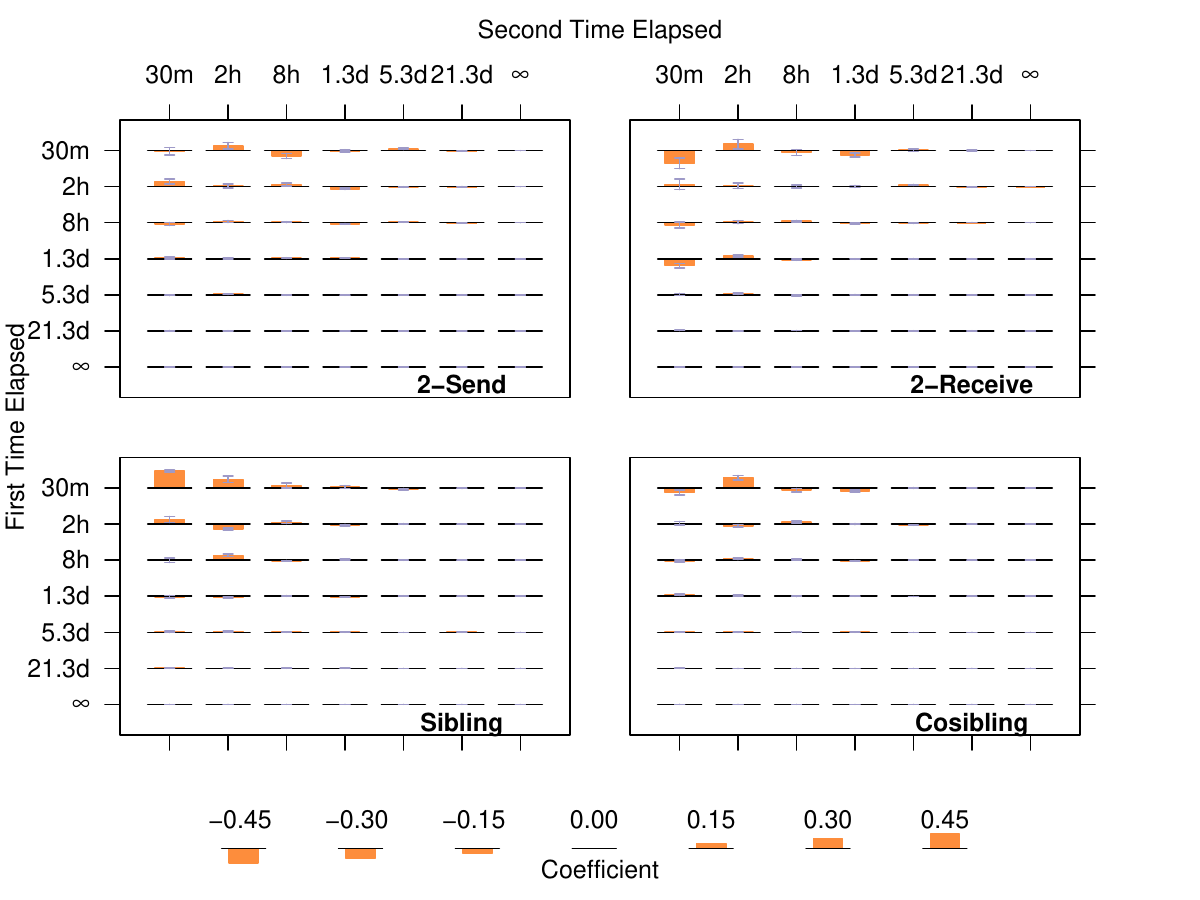}}
  \caption{Estimated coefficients for triadic effects, with standard errors}
  \label{F:enron-triad}
\end{figure}

The estimated coefficients for the recency-dependent covariates, shown in
Figs.~\ref{F:enron-dyad} and~\ref{F:enron-triad}, give a more complete
picture of network effects.  Firstly, we can see that dyadic effects
persist for over three weeks from the time a message is sent.  The decay of
the estimated coefficients is roughly exponential in the time elapsed,
corresponding to a super-exponential decay in the relative sending rate.  For 30
minutes after $i$ sends a message to $j$, our estimated model predicts that
the rate at which $i$ sends to $j$ will be multiplied by $\exp(1.11) \approx
3.05$, and the rate at which $j$ sends to $i$ will be multiplied by $\exp(1.85)
\approx 6.39$; then, between 30 minutes and 2 hours, the rates will be
multiplied by $\exp(0.51) \approx 1.67$ and $\exp(0.70) \approx 2.02$,
respectively; this proceeds similarly until after 21.3 days, when the rates
will be multiplied by $\exp(0.003) \approx 1.002$ and $\exp(0.002) \approx
1.002$.

Comparing the coefficients for $\textbf{send}_t^{(k)}$ with those of
$\textbf{receive}_t^{(k)}$ we see that the latter are higher for $k \leq 2$,
while the former are higher for $k > 2$.  The corresponding intuition is that if $A$ is
sending a message up to two hours after receiving a message from $B$, then $A$
is likely to respond to $B$, but after that, $A$ is more likely to send to an individual
whom $A$ e-mailed at the time of receiving $B$'s original message (provided
$B$ and this other individual are identical in all other respects).  The time window
during which reciprocation is more important than past habit is less than 8 hours.

From Fig.~\ref{F:enron-triad}, we can see that the triadic effects are in general less
pronounced and are much more short-lived than the dyadic effects.  About
86\% of the estimated coefficients are within $3$ standard errors of
$0$; even those that are significantly nonzero mostly lie between $-0.05$ and
$+0.05$.  The exceptions are the coefficients for
$\textbf{sibling}_t^{(1,1)}$ ($0.51$),
$\textbf{sibling}_t^{(2,2)}$ ($-0.14$),
$\textbf{sibling}_t^{(3,2)}$ ($0.15$),
$\textbf{cosibling}_t^{(1,2)}$ ($0.32$),
$\textbf{2-receive}_t^{(4,1)}$ ($-0.21$),
and
$\textbf{2-receive}_t^{(4,2)}$ ($0.09$).
We may interpret these coefficients as follows:

\begin{description}

\item[sibling] If $B$ sent $A$ and $C$ messages in the last 30 minutes or
between two and eight hours ago, then $A$ and $C$ are more likely to send
messages to each other; however, if $B$ sent $A$ and $C$ messages between 30
minutes and two hours ago, then $A$ and $C$ are less likely to send messages
to each other.

\item[cosibling] If $A$ sent a message to $B$ in the last 30 minutes, and $C$
sent a message to $B$ between 30 minutes and two hours ago, then $A$ will send
to $B$ at a higher rate.

\item[2-receive] If $A$ sent a message to $B$ in the last 30 minutes, and $B$
sent a message to $C$ between 8 hours and 32 hours ago, then $C$ will send
to $A$ at a lower rate; if, however, the message from $A$ to $B$ was sent
between 30 minutes and two hours ago, then $C$ will send to $A$ at a higher
rate.

\end{description}

Given the emphasis on transitivity in the networks literature, it may at first
seem disconcerting that most of the estimated coefficients for the
time-dependent triadic effects are found to be insignificant in this analysis.
However, one must bear in mind that, except for  messages sent to them
directly, individuals likely have no knowledge of their colleagues' e-mail
activities, and therefore there is no reason why this activity should directly
affect sending behaviors.  Any predictive power the triadic effects have,
then, must be due to correlation with exogenous factors.  In this light, it is
not surprising that the triadic effects are small and have small time
horizons.

The results above provide a detailed view of the
ways in which network effects can manifest themselves in data.
The supplementary material
contains comparative analyses based on
an actor-oriented model and an exponential random graph model.  (See \citet{snijders2010introduction} and \citet{anderson1999pstar}, respectively, for detailed surveys.) These analyses
further bolster our confidence in the results of this section.

\FloatBarrier

\section{Conclusion}\label{S:discussion}

Our analysis of the Enron corpus in Sections~\ref{S:enron-modeling}
and~\ref{S:strength-of-effects} above has demonstrated the ways in which
static and dynamic effects manifest themselves in e-mail communication
networks, and we expect similar conclusions to hold broadly for other types of
directed interaction data.  Relative to alternatives such as contingency table
analyses, actor-oriented network models, and exponential random graph models,
an advantage of our approach lies in its ability to model the given data
directly, rather than in an aggregated form.  We are able to adjust for
network effects to get more reliable estimates of homophily, and by using
continuous-time information we get precise quantification on the
time-dependent behavior of the network effects.

In this work, our focus has been on the coefficient vector $\beta$.  We have used partial likelihood for its estimation, enabling us to treat each sender-specific baseline intensity $\bar \lambda_t(i)$ as a nuisance parameter.  Were we to use the model for
prediction, we would need to estimate baseline intensities; this could be
done using a Nelson-Aalen estimator as in~\citet{andersen1993statistical}.

The foundation of our work is Cox's \citeyearpar{cox1972regression}
proportional intensity model and partial likelihood theory, tools
which he first introduced almost forty years ago and which have been
significantly developed since then \citep{cox1975partial,
fleming1991counting, andersen1993statistical, martinussen2006dynamic,
cook2007statistical}. These tools are used extensively in the context
of survival analysis, but require further development for use in
modeling interaction data. In this vein, we have extended the
associated theory in two directions: first, we have provided results
that are asymptotic in time rather than in the size of the population
under study; and second, we have shown that treating multicast
interactions via duplication leads to bias in the parameter estimates
(which can in turn be corrected in certain regimes).

We find the
proportional intensity model with time-varying covariates to be particularly
useful for modeling repeated directed interactions.  The model is
simple, flexible, and well established, and it facilitates investigation
into which traits and behaviors are predictive of interaction.

\section*{Acknowledgement}

We thank Joe Blitzstein, Susan Holmes, Art Owen, and Andrew Thomas for helpful
remarks and encouragement.  We benefited from many helpful comments by the
reviewers at JRSS, who brought the paper by \cite{butts2008relational} to our
attention and pushed us to expand the data analysis section.  Work supported
in part by the Army Research Office under PECASE Award W911NF-09-1-0555,
the Office of Naval Research under MURI Award 58153-MA-MUR, and the Royal Society
under a Wolfson Research Merit Award.

\appendix

\section{Implementation}\label{S:implementation}

To compute the maximum partial likelihood estimator, we use Newton's
method as described in \citet{boyd2004convex}.
This requires an efficient algorithm for
computing the gradient and Hessian of the log partial likelihood.  For
simplicity, we describe the case of strictly pairwise interactions with no
ties in the interaction times.  We use the notation from
Section~\ref{S:point-process-model}, with
the model from~\eqref{E:cox-intensity} and the partial likelihood
from~\eqref{E:log-pl}.  Recall that $x_t(i,j)$ is in $\reals^p$.
Assume that $|\mathcal{I}| = I$ and $|\mathcal{J}| = J$.

Suppose $(t_1, i_1, j_1), \ldots, (t_n, i_n, j_n)$ is the sequence of observed
interactions.  Set $n(i) = \#\{ i_m : i_m = i \}.$
The partial likelihood factors into a product of terms, one for each sender:
\begin{equation*}
    \mathit{PL}_t(\beta)
        =
        \,\,
        \prod_{i \in \mathcal{I}}
            \,\,
            \mathit{PL}_t(\beta, i),
    \qquad
    \mathit{PL}_t(\beta, i)
        =
        \!\!\!\!
        \prod_{\substack{t_m \leq t, \\ i_m = i}}
            \!\!\!
            \frac{w_{t_m} (\beta, i, j_m)}
                 {W_{t_m}(\beta, i)}.
\end{equation*}
This factorization allows us to compute $\log \mathit{PL}_t(\beta)$ and
its derivatives by computing the sender-specific terms in parallel and
then adding them together.

The gradient and Hessian of the sender-specific log partial likelihood
are respectively
\begin{subequations}
\begin{align}\label{eq:grad-MPLE}
    \nabla [ \log \mathit{PL}_t(\beta, i) ]
        & =
        \sum_{\substack{t_m \leq t, \\ i_m = i}}
            x_{t_m}\!(i,j_m)
            -\!\!
            \sum_{\substack{t_m \leq t, \\ i_m = i}}
                E_{t_m}(\beta, i) \\
    -\nabla^2 [ \log \mathit{PL}_t(\beta, i) ]
        & =
        \sum_{\substack{t_m \leq t, \\ i_m = i}}
          V_{t_m}(\beta, i),\label{eq:hess-MPLE}
\end{align}
\end{subequations}
where $E_{t}(\beta,i)$ and $V_{t}(\beta, i)$ are as defined in
~\eqref{E:log-pl-gradient} and~\eqref{E:log-pl-neg-hessian}.
When $x_t(i,j)$ is constant over time, sufficient statistics for $\beta$ imply that these formulae simplify.
Otherwise, computing the first two derivatives of
$\log \mathit{PL}_{t_n}(\beta)$
necessitates iterating over all messages, potentially requiring time
$\Oh(n \, J \, p^2)$.  For small- to medium-sized datasets, this is manageable,
but for large network datasets it can become
prohibitive.  In the sequel we show how to exploit sparsity to drastically
reduce the computation time.

\subsection{Initial values}

We will need to compute $W_0(\beta, i)$, $w_0(\beta, i, j)$, $E_0(\beta,
i)$, and $V_0(\beta, i)$ for all values of $i$ and $j$.  In the worst case,
doing so will take $\mathcal{O}(I \, J \, p^2)$.  However, often the senders
belong to a small number, $\bar I \ll I$ of groups such that if $i$ and $i'$
are in the same group, then the corresponding values of $W_0$, $\pi_0$, $E_0$,
and $V_0$ are the same, reducing the total complexity to $\mathcal{O}(\bar I
\, J \, p^2)$.  The remaining complexity estimates assume that the
initial values have all been pre-computed.

\subsection{Exploiting sparsity}

We first decompose $x$ into its static (non-time-varying) and dynamic parts as follows:
\begin{equation}\label{E:x-static-dynamic}
    x_t(i,j)
        = x_0(i,j) + \Delta x_t(i,j).
\end{equation}
Typically, we can quickly compute the dynamic part $\Delta x_t(i,j)$ at
each observed message time by incrementally updating it.  Further, $\Delta
x_t(i,j)$, is zero for most $(i,j)$ pairs---often $\Delta x_t(i,j)$ is zero
unless $i$ and $j$ have a common acquaintance or they have interacted in the
past.
For convenience, set $\mathcal{J}_0(i) = \mathcal{J}$.  Let
\begin{equation*}
    \mathcal{\bar J}(i)
        =
            \{
                j \in \mathcal{J} :
                \text{
                    $j \in \mathcal{J}_t(i)$ and $\Delta x_t(i,j) \neq 0$
                    for some $t$
                }
            \}
        \cup
            \{
                j \in \mathcal{J} :
                \text{
                    $j \notin \mathcal{J}_t(i)$
                    for some $t$
                }
            \}.
\end{equation*}
For fixed $t$ and $i$, assume that computing
$\Delta x_t(i,j)$ for all values of $j$ takes amortized time $\Oh(d \bar J)$.

Since $\mathcal{J}_0(i) = \mathcal{J}$, we have that
\begin{align*}
    w_{t}(\beta,i,j)
        &=
            w_{0}(\beta,i,j)
            \cdot
            \exp\{ \beta^\trans \Delta x_t(i,j) \}
            \cdot
            1\{j \in \mathcal{J}_t(i)\} \nonumber \\
        &= w_0(\beta, i, j) + \Delta w_t(i, j);
    \\
    W_{t}(\beta,i)
        &=
            W_{0}(\beta,i)
            +
            \sum_{j \in \mathcal{\bar J}(i)}
                \Delta w_t(i,j);
\intertext{where}
  \Delta w_t(i,j)
    &= w_0(\beta, i, j)
      [\exp\{ \beta^\trans \Delta x_t(i,j)\}
          1 \{ j \in \mathcal{J}_t(i)\}
       - 1];
\end{align*}
here we have used that $\Delta w_t(i,j)$ is zero unless
$j \in \mathcal{\bar J}(i)$.
Write
\[
  \pi_t(\beta, i, j) = \frac{w_t(\beta, i, j)}{W_t(\beta, i)};
\]
then, defining
\[
  \gamma_t(i) = \frac{W_0(\beta, i)}{W_t(\beta, i)},
  \qquad\qquad
  \Delta \pi_t(\beta, i, j) = \frac{\Delta w_t(\beta, i, j)}{W_t(\beta, i)},
\]
we can express $\pi_t(\beta, i, j)$ as follows:
\[
  \pi_t(\beta, i, j)
    = \gamma_t(i) \pi_0(\beta, i, j) + \Delta \pi_t(\beta, i, j).
\]
Moreover, given the initial values $W_0(\beta, i)$ and $w_0(\beta, i, j)$,
we can efficiently keep track of $\gamma_t(i)$ and
$\Delta \pi_t(\beta, i, j)$: for any $i$ and $t$, it takes amortized time
$\Oh(\bar J d p)$ to evaluate $\gamma_t(i)$ and all values of
$\Delta \pi_t(i,j)$ as $j$ varies.

\subsection{Computing the gradient}

In evaluating the gradient of the log partial likelihood as given
by~\eqref{eq:grad-MPLE}, the sum $\sum_m x_{t_m}(i, j_m)$ can be computed in
time $\Oh(n \, p)$, while the computationally expensive term is \( \sum_{m}
E_{t_m}(\beta, i_m).  \) In the sequel we show how to exploit sparsity in $x$
to reduce the associated computational overhead.

To simplify the notation, we suppress the dependence of all quantities on $\beta$ and $i$.
Consider $\pi_t$ and $\Delta \pi_t$ to be vectors of length $J$, and write
\[
    \pi_t = \gamma_t \pi_0 + \Delta \pi_t.
\]
Also, let $X_t = X_t(i)$ and $\Delta X_t = \Delta X_t(i)$ be the $J \times p$ matrices
whose $j$th rows are $x_t(i,j)$ and $\Delta x_t(i,j)$, respectively, so
that
\[
  X_t = X_0 + \Delta X_t.
\]
Using these expressions, we obtain
\[
  E_t = X_t^\trans \pi_t
      = \gamma_t E_0
      + X_0^\trans \Delta \pi_t
      + \Delta X_t^\trans \pi_t,
\]
and thus,
\[
  \sum_{\substack{m  \\ i_m = i}} E_{t_m} =
    \Big(\sum_{\substack{ m \\ i_m = i}} \gamma_{t_m}\Big) E_0
    +
    X_0^\trans \Big(\sum_{\substack{ m \\ i_m = i }} \Delta \pi_{t_m}\Big)
    +
    \sum_{\substack{m \\ i_m = i}} \Delta X_{t_m}^\trans \pi_{t_m}.
\]
Taking advantage of the sparsity in $\Delta X_t$ and $\Delta \pi_t$, computing
the three sums on the right hand side takes time $\Oh\big(n(i) \,
\bar J \, d \, p\big)$.  Once the sums are known, the multiplication
$\Big(\sum \gamma_{t_m}\Big) E_0$ takes time $\Oh(p)$, and the
multiplication $X_0^\trans \big( \sum \Delta \pi_{t_m}\big)$ takes time
$\Oh(\bar J \, p)$.  Thus, we can compute $\sum_{\substack{m \\ i_m = i}} E_{t_m}$ in time
$\Oh\big(n(i) \, \bar J \, d \, p\big)$.  Computing these terms separately for
each $i$ and then summing over all $i$ to get the total gradient requires time $\Oh(n
\, \bar J \, d \, p + I \, p)$.

\subsection{Computing the Hessian}

Computing the Hessian according to~\eqref{eq:hess-MPLE} proceeds similarly to the case of the
gradient.  We need to efficiently compute the sum $\sum_m V_{t_m}(\beta,
i_m)$; while a naive computation requires time $\Oh(n \, J \, p^2)$, this can be significantly improved by exploiting sparsity in $x_t(i,j)$.

To this end, define $\Pi_t(\beta, i)$ to be the $J \times J$ diagonal matrix with
$[\Pi_t(\beta, i)]_{jj} = \pi_t(\beta, i, j)$, and set
$\Delta \Pi_t(\beta, i) = \Pi_t(\beta, i) - \Pi_0(\beta, i)$.
Suppressing the dependence on $\beta$ and $i$, we have
\begin{align*}
  V_t
    &= X_t^\trans [\Pi_t - \pi_t \pi_t^\trans] X_t \\
\begin{split}
    &= X_0^\trans [\Pi_t - \pi_t \pi_t^\trans] X_0
      \;+\; \Delta X_t^\trans [\Pi_t - \pi_t \pi_t^\trans] X_0\\
      &\quad+\; X_0^\trans [\Pi_t - \pi_t \pi_t^\trans] \Delta X_t
      \;+\; \Delta X_t^\trans [\Pi_t - \pi_t \pi_t^\trans] \Delta X_t.
\end{split}
\end{align*}
The first of these terms reduces to
\begin{align*}
  X_0^\trans [ \Pi_t - \pi_t \pi_t^\trans] X_0
    &= \gamma_t V_0
    \;+\; \gamma_t (1 - \gamma_t) E_0 E_0^\trans
    \;-\; E_0 (\gamma_t \Delta \pi_t)^\trans X_0^\trans  \\
    &\quad-\; X_0 (\gamma_t \Delta \pi_t) E_0^\trans
    \;+\; X_0^\trans [ \Delta \Pi_t - \Delta \pi_t \Delta \pi_t^\trans ] X_0,
\end{align*}
and the second can be expressed as
\begin{align*}
  \Delta X_t^\trans [ \Pi_t - \pi_t \pi_t^\trans] X_0
    &= (\gamma_t \Delta X_t \pi_t) E_0^\trans
    \;+\; \Delta X_t^\trans [\Pi_t + \pi_t \Delta \pi_t^\trans] X_0.
\end{align*}
The third term is the transpose of the second; the fourth does not
simplify.

To compute the sum $\sum_{\substack{m \\ i_m = i}} V_{t_m}$, we only
accumulate sums of terms that change with time:
$\gamma_t$,
$\Delta \pi_t$,
$\gamma_t(1 - \gamma_t)$,
$\gamma_t \Delta \pi_t$,
$\Delta \pi_t \Delta \pi_t^\trans$,
$\gamma_t \Delta X_t \pi_t$,
$\Delta X_t^\trans [\Pi_t + \pi_t \Delta \pi_t^\trans]$,
and
$\Delta X_t^\trans [\Pi_t - \pi_t \pi_t^\trans] \Delta X_t$.
Doing so takes time $\Oh(\bar J \, d \, p^2)$ for each time increment.  As with
the gradient computation, we compute the sums separately for each $i$ and then
sum over all $i$, so that the total computation time is
$\Oh(n \, \bar J \, d \, p^2 + I \, p^2)$.

\subsection{Total computation time}

To perform one Newton step in maximization of the log partial likelihood of~\eqref{E:log-pl}, we must first compute the gradient and Hessian of the
log partial likelihood at the current value of $\beta$, and then compute the
inverse of the Hessian and its product with the gradient.  Once we have the
Hessian, computing its inverse takes time $\Oh(p^3)$.  Typically, it takes
$\Oh(1)$ Newton steps to compute the maximum of a convex function (the
constant is often below $30$).  The key factors in determining the
computation time using the factors laid out above are $\bar I$, $\bar J$, and
$d$:

\begin{itemize}

\item The value of $\bar I$ depends on the structure of $x_0(i,j)$.  Specifically,
$\bar I$ is equal to the number of
distinct values of the matrix $X_0(i)$ as $i$
varies.  For the Enron data, we have that $\bar I = 12$: each sender
belongs to one of $12$ groups determined by group (L/T/O), seniority (J/S),
and gender (F/M), and so the matrix $X_0(i)$ depends only on the group of $i$.

\item The value of $\bar J$ depends on the sparsity of $x_t(i,j)$.  If
$x_t(i,j)$ includes only dyadic network effects, then $\bar J$ will typically
be of size $\Oh(1)$ or $\Oh(J^\alpha)$ for a fractional value $\alpha$; when we add triadic effects, this size will typically
grow to at most $\Oh(J^{2 \alpha})$.

\item The value of $d$ depends on further structure in $x_t(i,j)$.  In our
implementation, $d = \Oh(1)$ for dyadic effects and $d = \Oh(\bar J)$ for
triadic effects.

\end{itemize}

The total computational cost per Newton step is thus
$\Oh(\bar I \, J \, p^2  + n \, \bar J \, d \, p^2 + I \, p^2 + p^3)$, with
the significance of this expression being that it is nearly linear
in $I$, $J$, and $n$.  Thus, the algorithm scales naturally to large
datasets.

\newcommand{\MPLEconsistencysection}{\ref{S:MPLE-consistency}}
\section{Results from Section~\protect\MPLEconsistencysection{}}
\label{S:MPLE-consistency-proofs}

\subsection{Proof of Theorem~\ref{T:score-fisher}}

Observe that the process $N_t(i,j)$ has compensator
\(
    \Lambda_t(i,j)
        =
            \int_0^t \lambda_s(i,j) \, ds;
\)
similarly, processes $N_t(i)$ and $N_t$ have compensators
$\Lambda_t(i) = \sum_{j \in \mathcal{J}} \Lambda_t(i,j)$
and $\Lambda_t = \sum_{i \in \mathcal{I}} \Lambda_t(i)$.  Correspondingly,
define local martingales $M_t(i,j) = N_t(i,j) - \Lambda_t(i,j)$,
$M_t(i) = N_t(i) - \Lambda_t(i)$, and $M_t = N_t - \Lambda_t$;
also define
\[
    H_t(i,j)
        =
        x_t(i,j) - E_t(\beta_0,i),
\]
where $E_t(\beta,i)$ is as defined in~\eqref{E:wt-expectation}.

As observed by \citet{andersen1982cox}, the score function
$U_t(\cdot)$ evaluated at $\beta_0$ has
a simple representation in terms of these processes:
\begin{align*}
    U_t(\beta_0)
        &=
        \sum_{i \in \mathcal{I}}
        \sum_{j \in \mathcal{J}}
        \int_0^t
            H_s(i,j) \, dN_s(i,j)
        =
        \sum_{i \in \mathcal{I}}
        \sum_{j \in \mathcal{J}}
        \int_0^t
            H_s(i,j) \, dM_s(i,j),
\end{align*}
since
\(
    \sum_{j \in \mathcal{J}}
    \int_0^t
        H_s(i,j) \,
        d\Lambda_s(i,j)
    =
    0.
\)
Since by Assumption A\ref{A:square-int}, $x$ is uniformly bounded, $H$
is as well.  Each term in the sum above is thus locally square integrable,
with predictable covariation
\begin{align*}
    \begin{split}
        \bigg\langle
            \int
                H_s(i,j) \, dM_s(i,j)
        &, \, \,
            \int
                H_s(i',j') \, dM_s(i',j')
        \bigg\rangle_t \\
        &=
            \int_0^t
                H_s(i,j) \otimes H_s(i',j') \,
                d\big\langle M(i,j), M(i',j')\big\rangle_s
    \end{split} \\
        &=
            \int_0^t
                \big[ H_s(i,j) \big]^{\otimes 2} \,
                d\Lambda_s(i,j)
            \cdot
            1\{ i = i', j = j' \}
\end{align*}
\citep[Thm.~2.4.3]{fleming1991counting}.  There exists a sequence
of stopping times localizing all $M(i,j)$ simultaneously, so $U(\beta_0)$ is
locally square integrable with predictable variation
\begin{align}
\begin{split}\label{E:score-compensator}
    \big\langle U(\beta_0) \big\rangle_t
        &=
            \sum_{i \in \mathcal{I}}
            \sum_{j \in \mathcal{J}}
            \int_0^t
                \big[ H_s(i,j) \big]^{\otimes 2} \,
                d\Lambda_s(i,j)
        =
            \sum_{i \in \mathcal{I}}
            \int_0^t
                V_s(\beta_0, i) \,
                d\Lambda_s(i).
\end{split}
\end{align}

Now we rescale time.  For each positive $n$ define a discretized time-scaled
version of the score that is right-continuous with limits from the left.
The process is defined for times $\alpha$ in $[0,1]$; between times in
$[\tfrac{k}{n}, \tfrac{k+1}{n})$, it takes the value $U_{t_k}$; i.e.,
\begin{equation}\label{E:score-time-scaled}
    \tilde U_{\alpha}^{(n)}(\beta)
        = U_{t_{\lfloor \alpha n \rfloor}}(\beta).
\end{equation}

Part~\textit{(\ref{I:score-part})}:
Lemma~\ref{L:adapted-martingale} shows that $\tilde U_\alpha^{(n)}(\beta_0)$
is a square-integrable martingale adapted to
\(
    \mathcal{\tilde F}^{(n)}_\alpha
        =
        \mathcal{F}_{t_{\lfloor \alpha n \rfloor}},
\)
the $\sigma$-algebra of events prior to
$t_{\lfloor \alpha n \rfloor}$.
Since it only depends on values at jump times, the
quadratic variation of $\tilde U^{(n)}(\beta_0)$ at time $\alpha$ is
equal to the quadratic variation of $U(\beta_0)$ at time
$t_{\lfloor \alpha n \rfloor}$.  Therefore, since quadratic and
predictable variation have the same limit when it exists
\citep[Prop.~1]{rebolledo1980central}, assumption
A\ref{A:integrated-cov-limit} implies that
\(
    \langle \frac{1}{\sqrt{n}} \tilde U^{(n)}(\beta_0) \rangle_\alpha
        \toP
            \Sigma_\alpha(\beta_0).
\)
Lemma~\ref{L:Lindeberg-condition} in turn verifies that
$\frac{1}{\sqrt{n}} \tilde U^{(n)}(\beta_0)$ satisfies a Lindeberg
condition necessary for the application of Rebolledo's \citeyearpar{rebolledo1980central} Martingale Central
Limit Theorem.  Thus the process converges in
distribution to a Gaussian process with covariance function
$\Sigma_\alpha(\beta_0)$ as claimed.

Part~\textit{(\ref{I:fisher-part})}:
Recalling $M_t(i) = N_t(i) - \Lambda_t(i)$,
combine~\eqref{E:log-pl-neg-hessian} and~\eqref{E:score-compensator}
to obtain the relation
\begin{equation}\label{E:var-estimate-relation}
            \sum_i
            \int_0^{t_{\lfloor \alpha n \rfloor}}
                V_s(\beta_0, i) \, dM_s(i)
        =
            I_{t_{\lfloor \alpha n \rfloor}}(\beta_0)
        -
            \big\langle \tilde U^{(n)}(\beta_0) \big\rangle_\alpha.
\end{equation}

When $\alpha \in [0, 1]$, a repeated application of the triangle
inequality to
\[
    \left\|
        \tfrac{1}{n} I_{t_{\lfloor \alpha n \rfloor}}(\hat \beta_n)
        -
        \tfrac{1}{n} \big(
            I_{t_{\lfloor \alpha n \rfloor}}(\beta_0)
            -
            I_{t_{\lfloor \alpha n \rfloor}}(\beta_0)
        \big)
        -
        \Sigma_{\alpha} (\beta_0)
    \right\|
\]
using the relation of~\eqref{E:var-estimate-relation} yields
\begin{multline*}
    \Big\|
        \tfrac{1}{n} I_{t_{\lfloor \alpha n \rfloor}}(\hat \beta_n)
        -
        \Sigma_{\alpha} (\beta_0)
    \Big\|
        \leq
        \Big\|
            \frac{1}{n}
            \sum_i
            \int_0^{t_{\lfloor \alpha n \rfloor}}
                \{
                    V_s(\hat \beta_n, i)
                    -
                    V_s(\beta_0, i)
                \} \, dN_s(i)
        \Big\| \\
        +
        \Big\|
            \frac{1}{n}
            \sum_i
            \int_0^{t_{\lfloor \alpha n \rfloor}}
                V_s(\beta_0, i) \, dM_s(i)
        \Big\| 
        +
        \Big\|
            \frac{1}{n}
            \sum_i
            \int_0^{t_{\lfloor \alpha n \rfloor}}
                V_s(\beta_0,i)
                \, d\Lambda_s(i)
            -
            \Sigma_{\alpha}(\beta_0)
        \Big\|.
\end{multline*}
We show that all three terms converge to zero in probability.
The first term above is uniformly bounded by
\(
    \sup_{n',i}
        \|
            V_{t_{n'}}(\hat \beta_n, i)
            -
            V_{t_{n'}}(\beta_0, i)
        \|,
\)
which converges to zero since $\hat \beta_n \toP \beta_0$ by
hypothesis of the theorem and $\{ V_{t_{n'}}(\cdot, i) \}$ is an
equicontinuous family by assumption~A\ref{A:var-equicont}.
Lemma~\ref{L:Lenglart} proves, as a consequence of
assumption~A\ref{A:message-times-finite} and Lenglart's \citeyearpar{lenglart1977relation}
Inequality, that the second term converges
to zero uniformly in $\alpha$.
The third term converges to zero by
assumption~A\ref{A:integrated-cov-limit}, thereby concluding the proof.

\subsection{Supporting lemmas for Theorem~\ref{T:score-fisher}}

\begin{lemma}\label{L:adapted-martingale}
Using the notation of Theorem~\ref{T:score-fisher}, under assumption
A\ref{A:square-int} the process $\tilde U_\alpha^{(n)}(\beta_0)$
from~\eqref{E:score-time-scaled} is a square-integrable martingale adapted to
\(
    \mathcal{\tilde F}^{(n)}_\alpha
        =
        \mathcal{F}_{t_{\lfloor \alpha n \rfloor}}.
\)
\end{lemma}

\begin{proof}
The conditional expectation property holds provided
\(
    \E[ U_{t_{n}}(\beta_0) \!\mid\! \mathcal{F}_{t_{n-1}} ]
        = U_{t_{n-1}}(\beta_0).
\)
Define $K = \sup_{t,i,j} \| x_{t}(i,j) \|$.
Note that $\|H_{t}(i,j)\| \leq 2 K$.  Thus,
\begin{align*}
    \| U_{t \wedge t_n} (\beta_0) \|
        &\leq
            2 K
            \big(
                N_{t \wedge t_n}
                +
                \Lambda_{t \wedge t_n}
            \big), \\
    \E \left[
        \sup_t
        \| U_{t \wedge t_n} (\beta_0) \|^2
    \right]
        &\leq
            8 \cdot \big(\E K^2\big)^{1/2} \cdot
            \big(
              \E N_{t_n}^2
              +
              \E \Lambda_{t_n}^2
            \big)^{1/2}.
\end{align*}
By assumption A\ref{A:square-int}, $\E K^2$ is finite, and by construction,
$N_{t_n}$ is bounded.  Since $N_{t \wedge t_n}$ is a counting process,
$\E \Lambda_{t_n}^2$ is finite, too
(this follows from results in Section 2.3 of
\citet{fleming1991counting}).  Thus, $U_{t \wedge t_n}(\beta_0)$
is uniformly integrable.  The Optional Sampling Theorem now applies to
give the conditional expectation property of $\tilde U^{(n)}(\beta_0)$.  For
square integrability, note
\(
    \sup_{1 \leq m \leq n}
    \E \| U_{t_m} \|^2
        \leq
        \E \left[
           \sup_t
           \| U_{t \wedge t_n} (\beta_0) \|^2
        \right].
\)
\end{proof}

\begin{lemma}\label{L:Lindeberg-condition}
Using the notation of Theorem~\ref{T:score-fisher}, under assumption
A\ref{A:square-int}, the Lindeberg condition for Rebolledo's \citeyearpar{rebolledo1980central} Central Limit
Theorem is satisfied: for any positive $\varepsilon$,
\[
    \frac{1}{n}
    \sum_{i,j}
    \int_{0}^{t_n}
        \| H_s(i,j) \|^2
        \, 1\{\| H_s(i,j) \| > \sqrt{n} \varepsilon \}
        \, d\Lambda_s(i,j)
        \toP
        0.
\]
\end{lemma}

\begin{proof}
With $K = \sup_{t,i,j} \| x_t(i,j)\|$ as above, the integral is bounded by
\(
    4 \, K^2 \, 1\{n^{-1/2} K > \varepsilon / 2\}
    \cdot
    \frac{\Lambda_{t_n}}{n}.
\)
Since $\E K^2 < \infty$ by assumption A\ref{A:square-int}, the first term
converges to zero in probability.  Since
$\E \Lambda_{t_n} = \E N_{t_n} = n$, the product of the two also
converges to zero in probability.  Thus, the Lindeberg condition is satisfied.
\end{proof}

\begin{lemma}\label{L:Lenglart}
Using the notation of Theorem~\ref{T:score-fisher}, under assumptions
A\ref{A:square-int}~and~A\ref{A:message-times-finite} we have that
\(
    \left\|
            \frac{1}{n}
            \sum_i
            \int_0^{t_{\lfloor \alpha n \rfloor}}
                V_s(\beta_0, i) \, dM_s(i)
    \right\|
        \toP 0
\)
uniformly in $\alpha$.
\end{lemma}

\begin{proof}
$\!$Lenglart's \citeyearpar{lenglart1977relation} Inequality and assumption~A\ref{A:message-times-finite} imply that
for any positive $\rho$ and $\delta$,$\!$
\begin{equation*}
    \mathbb{P}\Big\{
        \sup_{t \in [0,t_n]}
        \Big\|
            \frac{1}{n}
            \sum_{i}
            \int_{0}^{t}
                V_s(\beta_0, i) \, dM_s(i)
        \Big\|
        \geq \rho
    \Big\}
    \leq
    \frac{\delta}{\rho^2}
    +
    \mathbb{P}\Big\{
        \frac{1}{n^2}
        \sum_{i}
        \int_{0}^{t_n}
            \| V_s (\beta_0, i) \|^2
            \, d\Lambda_s(i)
        \geq
        \delta
    \Big\}.
\end{equation*}
(see \citet[Cor.~3.4.1]{fleming1991counting} for a related proof).  As in the proof of
Lemma~\ref{L:adapted-martingale}, set $K = \sup_{t,i,j} \| x_t(i,j) \|$.
The sum is bounded by $\frac{16 K^4}{n} \cdot \frac{\Lambda_{t_n}}{n}$.
Since $n^{-1/2} K^2 \toP 0$ by assumption A\ref{A:square-int} and
$\E \Lambda_{t_n} = n$, the right-hand side of the inequality converges to
$\frac{\delta}{\rho^2}$.  Since $\delta$ is arbitrary, the right-hand side
must converge to zero.
\end{proof}

\subsection{Proof of Theorem~\ref{T:consistency}}\label{S:proof-consistency}

We follow Haberman's \citeyearpar{haberman1977maximum} approach to proving
consistency, which relies on Kantorovich's \citeyearpar{kantorovich1952functional} analysis
of Newton's method.  \citet{tapia1971kantorovich} gives an elementary proof of the Kantorovich
Theorem.  We state a weak form of the result as a lemma.

\begin{lemma}[Kantorovich Theorem]\label{L:kantorovich}
    Let $P(x) = 0$ be a general system of nonlinear equations, where $P$ is
    a map between two Banach spaces.  Let $P'(x)$ denote the Jacobian
    (Fr\'echet differential) of $P$ at $x$, assumed to exist in $D_0$,
    a convex open neighborhood of $x_0$.
    Assume that
    \begin{enumerate}
        \item $\| [P'(x_0)]^{-1} \| \leq B$,
        \item $\| [P'(x_0)]^{-1} P(x_0) \| \leq \eta$,
        \item $\| P'(x) - P'(y) \| \leq K \| x - y \|$,\quad
            for all $x$ and $y$ in $D_0$,
    \end{enumerate}
    with $h = B K \eta \leq \tfrac{1}{2}$.

    Let $\Omega_\ast = \{ x : \| x - x_0 \| \leq 2 \eta \}$.
    If $\Omega_\ast \subset D_0$, then the Newton iterates,
    $x_{k+1} = x_k - [P'(x_k)]^{-1} P(x_k)$, are well defined, remain
    in $\Omega_\ast$, and converge to $x^\ast$ in $\Omega_\ast$ such
    that $P(x^\ast) = 0$.  In addition,
    \[
        \| x^\ast - x_k \|
            \leq
                \frac{\eta}{h}
                \frac{(2h)^{2^k}}{2^k},
        \qquad
        k = 0, 1, 2, \ldots.
    \]
\end{lemma}

\begin{proof}[Theorem~\ref{T:consistency}]

Set $U_t(\cdot)$ and $I_t(\cdot)$ to be the gradient and negative
Hessian of the log partial likelihood, as defined in
(\ref{E:log-pl-gradient}--\ref{E:log-pl-neg-hessian}).  Since
$I_t(\beta)$ is a sum of rank-one matrices with positive weights,
it is positive semi-definite, and $\log \mathit{PL}_t(\cdot)$ is
a concave function.  By the assumption that the smallest
eigenvalue of $\Sigma_1(\cdot)$ is bounded away from zero in a neighborhood
of $\beta_0$, for $n$ sufficiently large, if $\log \mathit{PL}_t(\cdot)$ has a
local maximum in that neighborhood then it must be the unique global maximum.

We find the local maximum by applying Newton's method to the
gradient of $\tfrac{1}{n} \log \mathit{PL}_{t_n}(\cdot)$, taking
$\beta_0$ as the initial iterate.  Define
\(
   Z_n = -[\tfrac{1}{n} I_{t_n}(\beta_0)]^{-1} [ \tfrac{1}{n} U_{t_n}(\beta_0)].
\)
The first Newton iterate, $\beta_{n,1}$, is equal to $\beta_0 - Z_n$.
Part (b) of Theorem~\ref{T:score-fisher} and the
assumptions of the theorem imply $[\tfrac{1}{n} I_{t_n}(\beta_0)]^{-1}$
exists for $n$ large enough, so that $Z_n$ is well-defined.
Moreover, Part (a) of Theorem~\ref{T:score-fisher} and Slutsky's Theorem imply
$Z_n \toP 0$ and
$\sqrt{n} \, Z_n \tod \Normal(0,\, [\Sigma_1(\beta_0)]^{-1})$.

Now we may apply Kantorovich's Theorem to bound $\| \hat \beta_n - \beta_0 \|$
and $\| \hat \beta_n - \beta_{n,1} \|$ as follows.  By assumption, there exists
a neighborhood of $\beta_0$, say $D_0$, and finite $K$ and $B$, such that
\(
    \|
        \frac{1}{n} I_{t_n} (\beta)
        -
        \frac{1}{n} I_{t_n}(\beta')
    \|
    \leq
    K
    \|
        \beta
        -
        \beta'
    \|
\)
and
\(
    \| \frac{1}{n} [I_{t_n}(\beta_0)]^{-1} \| \leq B
\)
for $\beta, \beta' \in D_0$.
Define $\eta_n = \| Z_n \|$ and $h_n = B K \eta_n$, noting that $h_n$ and
$\eta_n$ are size $\OhP(n^{-1/2})$.  Thus, for $n$ large enough,
\begin{enumerate}
    \item $\| \hat \beta_n - \beta_0 \| \leq 2 \, \eta_n \toP 0$,
    \item
        \(
            \sqrt{n} \, \| \hat \beta_n - (\beta_0 - Z_n) \|
            \leq
            2 \sqrt{n} \, \eta_n \, h_n
            \toP 0.
        \)
\end{enumerate}
Thus, $\hat \beta_n \toP \beta_0$, and $\sqrt{n} (\hat \beta_n - \beta_0)$
and $\sqrt{n} \, Z_n$ converge weakly to the same limit.
\end{proof}

\newcommand{\multiplereceiverssection}{\ref{S:multiple-receivers}}
\section{Results from Section~\protect\multiplereceiverssection{}}
\label{S:multiple-recipient-proofs}

\subsection{Proof of Theorem~\ref{T:log-pl-multiple-approx-error}}

\begin{proof}[Theorem~\ref{T:log-pl-multiple-approx-error}]

When $J \subseteq \mathcal{J}_t(i)$, set
\(
    X_t(i, J) = \sum_{j \in J} x_t(i,j)
\)
and
\(
    w_t(\beta, i, J)
        = \exp\{ \beta^\trans X_t(i,J) \}.
\)
As a slight abuse of notation, when $j$ is an element of $\mathcal{J}_t(i)$,
take ``$w_t(\beta, i, j)$'' to mean $w_t(\beta, i, \{j\})$.  Define weights
\[
    W_t(\beta, i; L)
        = \sum_{\substack{J \subseteq \mathcal{J}_t(i), \\
                          |J| = L}}
              w_t(\beta,i,J),
  \qquad
    \widetilde W_t(\beta, i; L)
        =
            \Big[ \sum_{j \in \mathcal{J}_t(i)} w_t(\beta, i, j) \Big]^L,
\]
and note that the approximation error in
$\log \widetilde{\mathit{PL}}_t(\beta)$ comes from replacing
$W$ with $\widetilde W$.

The gradients of the weights are
\begin{subequations}
\begin{gather*}
    E_t(\beta, i; L)
        = \nabla \big[ \log W_t(\beta, i; L) \big]
        =
        \frac{1}{W_t(\beta, i; L)}
        \sum_{\substack{J \subseteq \mathcal{J}_t(i), \\
                        |J| = L}}\!
            w_t(i,J)
            \,
            X_t(i,J), \\
    \widetilde E_t(\beta, i; L)
        = \nabla \big[ \log \widetilde W_t(\beta, i; L) \big]
        =
        L
        \cdot
        \frac{
            \sum_{j \in \mathcal{J}_t(i)}
                w_t(\beta, i, j) \, x_t(i,j)
        }{
            \sum_{j \in \mathcal{J}_t(i)}
                w_t(\beta, i, j)
        }.
\end{gather*}
\end{subequations}
The second is the expectation of $\sum_{l = 1}^L x_t(i, j_l)$ when
$j_1, \ldots, j_L$ are drawn independently and identically from
$\mathcal{J}_t(i)$ with weights $w_t(\beta, i, \cdot)$; the first is the same
expectation, conditional on the event that $j_1, \ldots, j_L$ are all unique.
Let $\tilde{\mathbb{P}}_{t,\beta,i;L}$ and $\mathbb{P}_{t,\beta,i;L}$
denote the two probability laws for $j_1, \ldots, j_L$, and let
$\tilde{\mathbb{E}}_{t,\beta,i;L}$ and $\mathbb{E}_{t,\beta,i;L}$ denote
expectations with respect to them, so that
\(
    E_t(\beta,i;L)
        =
        \mathbb{E}_{t,\beta,i;L} \big[ \sum_{l=1}^L x_t(i,j_l)\big]
\)
and
\(
    \widetilde E_t(\beta,i;L)
    =
    \tilde{\mathbb{E}}_{t,\beta,i;L} \big[ \sum_{l=1}^L x_t(i,j_l)\big].
\)

The bound on
\(
    \nabla [\log \mathit{PL}_{t_n}(\beta) ]
    -
    \nabla [\log \widetilde{\mathit{PL}}_{t_n}(\beta) ]
\)
derives from a bound on
\(
    E_t(\beta,i;L)
    -
    \widetilde E_t(\beta,i;L).
\)
Write
\[
    E_{t}(\beta, i; L) - \widetilde{E}_t(\beta, i; L)
        =
        \mathbb{E}_{t,\beta,i;L}
            \Big[ \sum_{l=1}^L x_t(i,j_l) \Big]
        -
        \widetilde{\mathbb{E}}_{t,\beta,i;L}
            \Big[ \sum_{l=1}^L x_t(i,j_l) \Big].
\]
We define probability law $\mathbb{P}^\ast_{t,\beta,i;L}$ and
associated random variables $j_1, \ldots, j_L$ and
$\tilde \jmath_1, \ldots, \tilde \jmath_L$, such that marginally
$j_1, \ldots, j_L$ are distributed according to $\mathbb{P}_{t,\beta,i;L}$
and $\tilde \jmath_1, \ldots, \tilde \jmath_L$ are distributed according
to $\tilde{\mathbb{P}}_{t,\beta,i;L}$, but the variables are coupled to have
nontrivial chance of agreeing.  Then,
\begin{align*}
    \Big\| E_{t}(\beta, i; L) - \widetilde{E}_t(\beta, i; L) \Big\|
        &=
            \Big\|
            \mathbb{E}_{t,\beta,i;L}^\ast
            \Big[
                \sum_{l=1}^L x_t(i,j_l)
                -
                \sum_{l=1}^L x_t(i, \tilde \jmath_l)
            \Big]
            \Big\| \\
        &\leq
            2 L
            \cdot
            \Big[
                \sup_{j \in \mathcal{J}_t(i)}
                \| x_t(i,j) \|
            \Big]
            \cdot
            \mathbb{P}^\ast_{t,\beta,i;L}
            \Big\{
                (j_1, \ldots, j_L)
                    \neq
                    (\tilde \jmath_1, \ldots, \tilde \jmath_L)
            \Big\}
\end{align*}
The coupling is as follows:
\begin{enumerate}
    \item Draw $(\tilde \jmath_1, \ldots, \tilde \jmath_L)$ according to
        $\tilde{\mathbb{P}}_{t,\beta,i;L}$.
    \item If $(\tilde \jmath_1, \ldots, \tilde \jmath_L)$ are all unique,
        set $(j_1, \ldots, j_L) = (\tilde \jmath_1, \ldots, \tilde \jmath_L)$,
        otherwise draw $(j_1, \ldots, j_L)$ independently according to
        $\mathbb{P}_{t,\beta,i;L}$.
\end{enumerate}
With $K = \sup_{j \in \mathcal{J}_t(i)} \| x_t(i,j) \|$,
Lemma~\ref{L:coupling-prob-bound} shows
\[
    \mathbb{P}^\ast_{t,\beta,i;L}
    \Big\{
        (j_1, \ldots, j_L)
            \neq
            (\tilde \jmath_1, \ldots, \tilde \jmath_L)
    \Big\}
        \leq
        \binom{L}{2}
        \cdot
        \frac{\exp\{4 K \, \| \beta \|\}}{| \mathcal{J}_t(i) |}.
\]
The resulting bound on
\(
    \|
    \nabla [\log \mathit{PL}_{t}(\beta) ]
    -
    \nabla [\log \widetilde{\mathit{PL}}_{t}(\beta) ]
    \|
\)
now follows by expressing
\[
    \nabla \big[ \log \widetilde{\mathit{PL}}_t(\beta) \big]
    -
    \nabla \big[ \log \mathit{PL}_t(\beta )\big]
        =
        \sum_{t_m \leq t}
            E_{t_m}\!(\beta, i_m; |J_m|)
            -
            \widetilde{E}_{t_m}\!(\beta, i_m; |J_m|).
\]
Using
\(
    \big\| E_{t}(\beta, i; L) - \widetilde{E}_t(\beta, i; L) \big\|
        \leq
        K \, L^2 \, (L - 1)
        \,
        \frac{\exp\{4 K \, \| \beta \|\}}{| \mathcal{J}_t(i) |},
\)
we get
\[
    \Big\|
        \nabla\big[ \log \widetilde{\mathit{PL}}_t( \beta ) \big]
        -
        \nabla\big[ \log {\mathit{PL}}_t( \beta ) \big]
    \Big\| \\
        \leq
            K
            \exp\{4 K \| \beta \|\}
            \cdot
            \sum_{t_m \leq t}
                \frac{|J_m|^2(|J_m| - 1)}{|\mathcal{J}_{t_m}(i_m)|}.
\]
We get the final bound for the gradients by replacing the numerators of the
summands with $\sup_m |J_m|$.

Using the same methods, Lemma~\ref{L:hessian-approx-bound} derives the bound
on the difference in Hessians.
\end{proof}

\subsection{Supporting lemmas for Theorem~\ref{T:log-pl-multiple-approx-error}}

\begin{lemma}\label{L:coupling-prob-bound}
Using the notation and assumptions of
Theorem~\ref{T:log-pl-multiple-approx-error},
\[
    \mathbb{P}^\ast_{t,\beta,i;L}
    \Big\{
        (j_1, \ldots, j_L)
            \neq
            (\tilde \jmath_1, \ldots, \tilde \jmath_L)
    \Big\}
        \leq
        \binom{L}{2}
        \cdot
        \frac{\exp\{4 K \, \| \beta \|\}}{| \mathcal{J}_t(i) |},
\]
where $K = \sup_t \| x_t(i,j) \|$.
\end{lemma}
\begin{proof}
The left hand side is bounded by the probability that the samples
$\tilde \jmath_1, \dotsc, \tilde \jmath_L$ are all
unique, which can be bounded by
\begin{align*}
        \sum_{k < l}
            \mathbb{P}^\ast_{t,\beta,i;L}
            \{
                \tilde \jmath_k = \tilde \jmath_l
            \}
        =
            \binom{L}{2}
            \sum_{j \in \mathcal{J}_t(i)}
                \Big[
                    \frac{
                        w_t(\beta, i, j)
                    }{
                        \sum_{j' \in \mathcal{J}_t(i)} w_t(\beta, i, j')
                    }
                \Big]^2.
\end{align*}
Note
\(
    \exp\{-K \, \| \beta \|\}
        \leq w_t(\beta,i,j)
        \leq \exp\{K \| \, \beta \|\},
\)
so that
\[
    \sum_{j \in \mathcal{J}_t(i)}
        \Big[
            \frac{
                w_t(\beta, i, j)
            }{
                \sum_{j' \in \mathcal{J}_t(i)} w_t(\beta, i, j')
            }
        \Big]^2
    \leq
        \frac{\exp\{4 K \, \| \beta \|\}}{| \mathcal{J}_t(i) |}.
    \qedhere
\]
\end{proof}

\begin{lemma}\label{L:hessian-approx-bound}
Using the notation and assumptions of
Theorem~\ref{T:log-pl-multiple-approx-error},
\begin{equation*}
    \Big\|
        \nabla^2\big[ \log \widetilde{\mathit{PL}}_t(\beta) \big]
        -
        \nabla^2\big[ \log \mathit{PL}_t(\beta) \big]
    \Big\|
        \leq
            2K^2
            \exp\{4 K \| \beta \|\}
            \cdot
            \sum_{t_m \leq t}
                \frac{|J_m|^3 \, (|J_m| - 1)}{|\mathcal{J}_{t_m}(i_m)|}.
\end{equation*}
\end{lemma}
\begin{proof}
The argument is similar to the bound on the difference in gradients in
the proof of Theorem~\ref{T:log-pl-multiple-approx-error}.  The Hessians of
the weights are
\begin{align*}
    \begin{split}
    V_t(\beta,i;L)
        &=
        \nabla^2 \big[  \log W_t(\beta, i; L) \big]
        =
        \frac{1}{W_t(\beta,i;L)}
        \sum_{\substack{J \subseteq \mathcal{J}_t(i), \\
                        |J| = L}}
            w_t(\beta,i,J)
            \Big[
                X_t(i,J)
                -
                E_t(\beta,i;L)
            \Big]^{\otimes 2},
    \end{split} \\
    \begin{split}
    \widetilde V_t(\beta,i;L)
        &=
        \nabla^2 \big[ \log \widetilde W_t(\beta,i;L) \big]
        =
        L
        \cdot
        \frac{
            \sum_{j \in \mathcal{J}_t(i)}
                w_t(\beta, i, j)
                \Big[ x_t(i,j) - \tfrac{1}{L} \widetilde E_t(\beta,i;L) \Big]^{\otimes 2}
        }{
            \sum_{j \in \mathcal{J}_t(i)}
                w_t(\beta, i, j)
        }.
    \end{split}
\end{align*}
The first is the covariance matrix of $\sum_{l=1}^L x_t(i,j_l)$ under
$\mathbb{P}_{t,\beta,i;L}$; the second is the covariance matrix of the same
quantity under $\tilde{\mathbb{P}}_{t,\beta,i;L}$.
The result follows in the same manner as in the proof of
Theorem~\ref{T:log-pl-multiple-approx-error}.
The relevant intermediate bound is
\[
    \Big\| V_{t}(\beta, i; L) - \widetilde{V}_t(\beta, i; L) \Big\|
        \leq
        2\, K^2 \, L^3 \, (L - 1)
        \,
        \frac{\exp\{4 K \, \| \beta \|\}}{| \mathcal{J}_t(i) |}.
    \qedhere
\]
\end{proof}

\subsection{Proof of Theorem~\ref{T:mple-approx-error}}

\begin{proof}[Theorem~\ref{T:mple-approx-error}]

We know that Newton's method applied to
$\tfrac{1}{n}\log \widetilde{\mathit{PL}}_{t_n}(\cdot)$ converges to
$\tilde \beta_n$ after sufficiently many iterations.  We employ $\hat \beta_n$ as the
initial iterate and use the Kantorovich Theorem (Lemma~\ref{L:kantorovich})
to bound $\|\tilde \beta_n - \hat \beta_n\|$.

In the notation of the lemma, $P(\cdot)$ is the gradient of
$\tfrac{1}{n} \log \widetilde{\mathit{PL}}_{t_n}(\cdot)$ and
$P'(\cdot)$ is its Hessian.
The conditions of Theorem~\ref{T:mple-approx-error} imply assumptions
(a) and (c) hold uniformly in $n$ for some finite $B$ and $K$.
Set
\[
    \eta_n =
    \Big\|
        \Big[
            \nabla^2\big[
                \tfrac{1}{n}
                \log \widetilde{\mathit{PL}}_{t_n}(\hat \beta_{n})
            \big]
        \Big]^{-1}
        \Big[
            \nabla\big[
                \tfrac{1}{n}
                \log \widetilde{\mathit{PL}}_{t_n}(\hat \beta_{n})
            \big]
        \Big]
    \Big\|
\]
and set $h_n = B K \eta_n$.
Since $\nabla\big[\log {\mathit{PL}}_{t_n}(\hat \beta_{n})\big] = 0$,
Theorem~\ref{T:log-pl-multiple-approx-error} and the boundedness of the
inverse Hessian imply $\eta_n = \OhP(G_n/n)$.  Therefore, for $n$
large enough,
\[
    \| \tilde \beta_n - \hat \beta_n \|
        \leq \frac{\eta_n}{h}\frac{(2h)^{2^0}}{2^0}
        = 2 \eta_n
        = \OhP(G_n/n).
    \qedhere
\]
\end{proof}



\section*{Supplementary material}

# Point process modeling for directed interaction networks: Supplementary material

Patrick O. Perry

*Stern School of Business, New York University, USA*

Patrick J. Wolfe

*Department of Statistical Science, University College London, UK*

## 1. A comparative analysis based on contingency tables

Were we interested only in homophily, we might be tempted to forgo the proportional intensity model of (1) from the main text, and instead perform a contingency table analysis. However, as we now describe, such an analysis leads to very different conclusions about the predictive strength of homophily in our data.

For example, suppose that we are interested in testing for seniority-based homophily. Ignoring network effects and other dependency, we might model $\mathbb{P}\{i \to j \mid i\}$, the probability of employee $j$ being the recipient of a message given that employee $i$ is the sender, by way of a multinomial logit model:

$$\mathbb{P}\{i \to j \mid i\} \propto \exp\{\beta_J J(j) + \beta_{JJ} J(i) J(j)\}.$$

In this setting, Junior-Junior homophily would manifest in a positive value of $\beta_J + \beta_{JJ}$ and Senior-Senior homophily would manifest in a negative value of $\beta_J$.

Since there are $n_J = 82$ Junior executives and $n_S = 74$ Senior executives, and since the sender and receiver of a message are distinct, we have that

$$\mathbb{P}\{i \to j \mid i, J(i) = 1\} = \frac{e^{(\beta_J + \beta_{JJ})J(j)}}{(n_J - 1)e^{\beta_J + \beta_{JJ}} + n_S}$$

$$\mathbb{P}\{i \to j \mid i, J(i) = 0\}, = \frac{e^{\beta_J J(j)}}{n_J e^{\beta_J} + (n_S - 1)}.$$

In turn, we compute the corresponding maximum likelihood coefficient estimates using the entries of a $2 \times 2$ table that counts the number of messages exchanged between each group:

|        | Receiver |        |
|--------|----------|--------|
| Sender | Junior   | Senior |
| Junior | 7972     | 5833   |
| Senior | 3977     | 14479  |

The resultant estimates are $\hat{\beta}_J = -1.4$ and $\hat{\beta}_{JJ} = 1.6$, with (Wald) standard errors of about 0.02 for each. Indeed, these are exactly the estimated coefficients we would obtain if we were to use

*Address for correspondence:* Patrick O. Perry, Information, Operations, and Management Sciences Department, Stern School of Business, New York University, 44 West 4th St, New York, NY 10012, USA
E-mail: pperry@stern.nyu.edu

|        | Receiver |         |         |         |
| ------ | -------- | ------- | ------- | ------- |
| Sender | L        | T       | J       | F       |
| 1      | -1.48    | -1.74   | -1.83   | -0.25   |
|        | (0.04)   | (0.03)  | (0.03)  | (0.03)  |
| L      | 3.70     | 0.48    | 0.23    | -0.22   |
|        | (0.04)   | (0.05)  | (0.03)  | (0.03)  |
| T      | 1.06     | 1.92    | 1.11    | -0.21   |
|        | (0.06)   | (0.04)  | (0.04)  | (0.04)  |
| J      | -0.12    | 1.36    | 1.70    | 0.25    |
|        | (0.04)   | (0.04)  | (0.03)  | (0.03)  |
| F      | 0.87     | -0.58   | 0.07    | 0.83    |
|        | (0.04)   | (0.04)  | (0.03)  | (0.03)  |

**Fig. 1.** Estimated coefficients and standard errors for the contingency table-based analysis of Section 1; dark coefficients are significant (via Wald test) at level $10^{-3}$.

the proportional intensity model $\lambda_t(i,j) = \bar{\lambda}_t(i) \exp\{\beta_J J(j) + \beta_{JJ} J(i) J(j)\}$, a result that holds more generally for non-time-varying covariates.

One problem with this analysis is that we have marginalized over the other covariates (Gender and Department), potentially introducing a Simpson's paradox. This issue is easily rectified, however, by introducing covariates for senders and receivers; Fig. 1 above shows the resulting coefficient estimates.

The far more important problem is that this analysis implicitly assumes each message to be independent and identically distributed, conditional on the sender of the message. This assumption is blatantly false: common sense tells us that if Junior $A$ sends a message to Junior $B$, then the next time $B$ sends a message, $B$ is more likely to choose $A$ as a recipient. Any homophily effect present in these interaction data is thus likely to be exaggerated by reciprocation and other network effects. Indeed, comparing the contingency table-based estimates in Fig. 1 above with the estimates from the proportional intensity model with network effects in Fig. 7 from the main text, we can see that the coefficient estimates are much higher when we don't adjust for network effects. Thus even in cases where network effects themselves are not the object of primary interest, it is important to account for them when making inferences about the predictive strength of other covariates.

## 2. Comparative analyses using actor-oriented and exponential random graph models

A number of dynamic network models exist in the literature, including Hanneke, Fu, and Xing's (2010) exponential random graph model with time-varying coefficients, and Kolar, Song, Ahmed, and Xing's (2010) time-varying stochastic block model. Alternative approaches explicitly based on point processes, but excluding network effects and other covariate information, include that of Malmgren et al. (2009), who model activity at the level of the individual using a hidden Markov model, and of Heard et al. (2010), who work at the level of the dyad, assuming a piecewise-constant interaction rate. The closest match to our approach is given by the actor-oriented model of Snijders (2001, 2005), which we now detail in Section 2.1 below. Then, in Section 2.2, we provide a comparison to a static network analysis based on an exponential random graph model.

*2.1. Actor-oriented model analysis*

The actor-oriented model is designed for a sequence of snapshots $G_1, G_2, \ldots$, of network activity, where each $G_t$ is an $I \times J$ binary matrix representing pairwise connectivity between actors at time $t$. The model is best suited for ties that persist in time, not instantaneous events; it treats the sequence of networks as a first-order Markov chain, with the distribution of $G_t$ determined by $G_{t-1}$. Actors are assumed to change their ties between times $t-1$ and $t$ to maximize a stochastic utility function that depends on characteristics of the overall network. Essentially, given that the network is in state $G$, and that actor $i$ is allowed to make a change, he will change his link to actor $j$ according to

$$p(j|i, G) \propto \exp\Big\{ \sum_{s=1}^{S} \beta_s T_s\big(G(i \rightsquigarrow j)\big) + \sum_{s=1}^{S'} \beta'_s T'_s(G \setminus \{i \to j\}) \Big\},$$

where $T_s$ and $T'_s$ are network statistics; $\beta_s$ and $\beta'_s$ are unknown coefficients; $G(i \rightsquigarrow j)$ denotes the network obtained either by adding link $i \to j$ if it is absent or removing link $i \to j$ if it is present; $G \setminus \{i \to j\}$ denotes the network obtained by removing $i \to j$ if it is present. Additionally, the rate at which actor $i$ changes ties between observation times $t-1$ and $t$ is given by $\lambda(i)$, specified via another parametric model. (See Snijders et al. (2010) for a more thorough introduction.) The change probability function $p(j|i, G)$ plays a similar role to the multiplier $\exp\{x_t(i,j)^\mathrm{T}\beta\}$ in the proportional intensity model from the main text, and the change rate function $\lambda(i)$ plays a similar role to the baseline intensity $\bar{\lambda}_t(i)$.

For purposes of comparison, we specified a change probability model $p(j|i, G)$ with network statistics analogous to those used in Section 5.2 from the main text, and then we estimated coefficient sets $\beta$ and $\beta'$ analogous to the coefficients in the proportional intensity model. We used the `RSiena` package (Ripley and Snijders, 2011) to specify and fit the actor-oriented model after binning the Enron e-mail interaction data at regular intervals to obtain network snapshots $G_1$, $G_2$, and $G_3$. Here, $G_t(i,j) = 1$ if message $i \to j$ was observed in period $t$, and $G_t(i,j) = 0$ otherwise; periods 1–3 correspond to consecutive four-month periods in the year 2001. The subset we looked at contains 60% of the messages in the Enron corpus. We chose this particular subset and temporal resolution partially for computational reasons, but also to make the network change statistics (Jaccard coefficients) within the range recommended by `RSiena` (near 0.3). Approximately 2 hours' time was required to fit the model.

We included the following terms, chosen to mimic the covariates detailed in Section 5.2 from the main text:

(a) Outdegree/density (`out`). This statistic counts the number of outgoing ties; it is analogous to our $\bar{\lambda}_t(i)$, except that the rate is the same for each sender.

(b) Group-level edge effects (`traits`). One covariate is included for each identifiable first-order interaction of the form $X(i)Y(j)$, where $i$ is the sender and $j$ is the receiver; the covariate counts the number of edges $i \to j$ with $X(i)Y(j) = 1$. These effects correspond to the group-level effects in our model.

(c) Outdegree/density endowment (`outendow`). This statistic counts the number of deleted outgoing ties; it corresponds to the negative of the **send** term in our model.

(d) Reciprocity (`recip`). This statistic counts the number of reciprocal ties; i.e., edge sets of the form $\{i \to j, j \to i\}$. It corresponds to the **receive** term in our model.

(e) 3-cycles (`3cycle`). This statistic counts the number of cyclic triples; i.e., edge sets of the form $\{h \to i, i \to j, j \to h\}$. It corresponds to the **2-receive** term in our model.

|        | Receiver |         |         |         |
|--------|----------|---------|---------|---------|
| Sender | L        | T       | J       | F       |
| 1      | -0.65    | -0.10   | 0.13    | 0.21    |
|        | (0.12)   | (0.07)  | (0.07)  | (0.09)  |
| L      | **0.62** | -0.28   | -0.11   | -0.16   |
|        | (0.13)   | (0.10)  | (0.09)  | (0.11)  |
| T      | 0.46     | **0.45**| -0.15   | -0.46   |
|        | (0.16)   | (0.06)  | (0.07)  | (0.10)  |
| J      | 0.00     | 0.15    | **0.10**| -0.21   |
|        | (0.11)   | (0.06)  | (0.06)  | (0.09)  |
| F      | 0.19     | 0.33    | 0.08    | **0.13**|
|        | (0.12)   | (0.06)  | (0.07)  | (0.08)  |

(a) Trait effects (`traits`)

| Variate     | out    | outendow | recip  | 3cycle | ttriple |
|-------------|--------|----------|--------|--------|---------|
| Coefficient | -1.94  | -0.94    | 2.02   | -0.26  | 0.30    |
| (SE)        | (0.02) | (0.01)   | (0.06) | (0.03) | (0.01)  |

(b) Network effects

**Fig. 2.** Estimated effects for the actor-oriented model of Section 2.1

(f) Transitive triplets (`ttriple`). This statistic counts the number of transitive triples; i.e., edge sets of the form $\{h \to i, i \to h, h \to j\}$. It corresponds to the **2-send**, **sibling**, and **cosibling** terms in our model.

Note that after binning the interaction counts to form network snapshots as required by the actor-oriented model, it is impossible to separate the **2-send**, **sibling**, and **cosibling** effects. Further, the first-order Markov nature of the model restricts our ability to quantify the time decay of the dynamic network effects.

Figure 2 above shows the fitted coefficients for the actor-oriented model. We can see that the estimated network effects agree qualitatively with those in Fig. 8 from the main text, as outdegree/density endowment has a negative coefficient while reciprocity and transitive triplets have positive coefficients. Further, the dyadic coefficients are larger than the triadic coefficients. A discrepancy between the two models is that **2-receive** had a negligible effect in the proportional intensity model, while its analogue (`3cycle`) had a small negative effect in the actor-oriented model. One possible explanation for this discrepancy is that treating all ties as binary forces a negative bias in otherwise-unimportant network effects. With the limitation that ties are binary, when actor $i$ tries to maximize his stochastic utility, he is forbidden from reinforcing an existing $i \to j$ link; he is more likely to link to an actor $j'$ for which link $i \to j'$ is absent. To counteract this tendency, the coefficient of `3cycle` is forced to be negative.

### 2.2. Exponential random graph model analysis

As a final comparison, we fit an exponential random graph model to our data. This class of models—one of the more popular for estimating the importance of network effects—specifies a probability distribution for a single directed graph represented by a binary matrix $G$. It supposes

that $\mathbb{P}\{G = g\} \propto \exp\{\sum_{s=1}^{S} T_s(g)\}$, where $T_s(g)$ is a network statistic, for example the number of transitive triples in the graph. (See Anderson et al. (1999) for a detailed survey.)

To apply this form of model, we employed a reduction of our data to obtain a single directed graph $G$ as follows. Based on an "elbow" in the empirical cumulative distribution function of message counts $N_\infty(i,j)$ in our data, we chose a threshold of 10 sent messages and defined $G$ by

$$G(i,j) = 1\{N_\infty(i,j) \geq 10\}.$$

Next, as in our comparison to the actor-oriented model, we chose terms in the model to mimic the covariates from Section 5.2 of the main text. We used the `ergm` software package to fit the model (Handcock et al., 2011), based on a Markov chain Monte Carlo sample size of $10^5$ following a burn-in period of $10^6$ iterates. The covariates were as follows:

(a) Sender effects (`sender`). One covariate is included for each sender, measuring the outdegree of the sender. The corresponding coefficient plays the role of $\bar{\lambda}_t(i)$ in our model.

(b) Group-level edge effects (`edgecov`). One covariate is included for each identifiable first-order interaction of the form $X(i)Y(j)$, where $i$ is the sender and $j$ is the receiver; the covariate counts the number of edges $i \to j$ with $X(i)Y(j) = 1$. These effects correspond to the group-level effects in our model. We attempted to include second-order interactions as well, but were unable (for computational reasons) to fit the model with these terms.

(c) Mutuality (`mutual`). This statistic counts the number of mutual ties; i.e., edge sets of the form $\{i \to j, j \to i\}$, and corresponds to the **receive** term in our model.

(d) Cyclic triples (`ctriple`). This statistic counts the number of cyclic triples; i.e., edge sets of the form $\{h \to i, i \to j, j \to h\}$, and corresponds to the **2-receive** term in our model.

(e) Transitive triples (`ttriple`). This statistic counts the number of transitive triples; i.e., edge sets of the form $\{h \to i, i \to j, h \to j\}$. It corresponds to the **2-send**, **sibling**, and **cosibling** terms in our model.

Note that reducing the interaction data to a single directed graph has important modeling consequences. As with the case of the snapshot-based actor-oriented model detailed above, it is impossible to separate the **2-send**, **sibling**, and **cosibling** effects, and the inability to include second-order interactions between group-level effects precludes a direct comparison with the estimated group-level effects from the proportional intensity model. Further, for a single directed graph, there is no possibility of including a term corresponding to **send**, and it is impossible to quantify the time-dependence of the network effects.

Figure 3 overleaf shows the fitted coefficients for the exponential random graph model. As with the case of the actor-oriented model considered in Section 2.1 above, the estimated network effects agree qualitatively with those of Fig. 8 from the main text. Specifically, mutuality and transitive triples have positive effects, while cyclic triples have a negligible effect (in contrast to the case of Fig. 2 from Section 2.1 above).

| Sender | Receiver | | | |
|---|---|---|---|---|
| | L | T | J | F |
| 1 | -2.61 | -1.49 | -0.87 | -0.55 |
| L | 2.60 | 0.77 | -0.09 | 0.12 |
| T | 0.63 | 0.69 | -0.60 | -0.32 |
| J | 0.13 | 1.04 | 1.15 | 0.27 |
| F | 0.25 | 0.49 | 0.49 | 0.79 |

(a) Trait effects

| Variate | mutual | ctriple | ttriple |
|---|---|---|---|
| Coefficient | 4.49 | 0.15 | 0.42 |
| (SE) | (0.003) | (0.290) | (0.060) |

(b) Network effects

**Fig. 3.** Estimated coefficients for the exponential random graph model of Section 2.2. The model also included a term for each sender (not shown); furthermore, standard errors returned for the group-level edge effects were nonsensical, and so are not reported.



\end{document}